\documentstyle[psfig]{article}
\def\bitem{\par\smallskip\noindent\hangindent12pt\small}

\def\fm{\hbox{$.\!\!^{\rm m}$}}

\def\farcs{\hbox{$.\!\!^{\prime\prime}$}}

\def\phd0{\hbox{$\phantom{\hbox{.0}}$}}
\newcommand\aj{{AJ}}%
\newcommand\apj{{ApJ}}%
\newcommand\apjs{{ApJS}}%
\newcommand\apss{{Ap\&SS}}%
\newcommand\aap{{A\&A}}%
\newcommand\aaps{{A\&AS}}%
\newcommand\mnras{{MNRAS}}%
\newcommand\pasp{{PASP}}%
\newcommand\bain{{Bull.~Astron.~Inst.~Netherlands}}%
\newcommand{\Ha}{H$\alpha$ }
\newcommand{\Hza}{H$\alpha$}
\newcommand{\Hb}{H$\beta$ }
\newcommand{\Hzb}{H$\beta$}

\newcommand{\OIII}{[O~III]~$\lambda$5007 }
\newcommand{\OzIII}{[O~III]~$\lambda$5007}

\newcommand{\HST}{{\it Hubble Space Telescope }}

\newcommand{\ibvs}{IBVS}

\newcommand{\adap}{Annales d'Astrophysique}

\begin{document}
\markright{R.A. Downes et al.: Luminosities 
of [O~III] and Balmer lines in nova shells}

\title{\parbox{\textwidth}{ {\normalsize \sl 2001: The Journal of 
Astronomical Data 7, 1.   \hfill
\copyright{}~R.A. Downes et al.  }}\\ {\vspace{5mm} 
Luminosities of \OIII and Hydrogen Balmer lines in nova shells 
years and decades after outburst}}

\author{
Ronald A. Downes (1)\thanks{Visiting Astronomer, 
Kitt Peak National Observatory, National Optical
Astronomy Observatories, which is operated by the Association of Universities
for Research in Astronomy, Inc.\ under contract with the National Science
Foundation} ~and 
Hilmar W. Duerbeck (2)\thanks{Based on observations collected at the 
European Southern Observatory, La Silla, Chile} 
\\
{\small with the collaboration of}\\
Cath\'erine E. Delahodde (3)$^\dag$\\
\small (1) Space Telescope Science Institute, 3700 San Martin Drive, 
Baltimore, MD 21218, USA \\
\small (2) University of Brussels (VUB), Pleinlaan 2, 1050 Brussels,
           Belgium\\
\small (3) European Southern Observatory, A. de Cordova 3107, Santiago, Chile\\
\small (now at Institut d'Astrophysique de Marseille, Traverse du Siphon,\\
\small F-13376 Marseille Cedex 12, France)}
\date{\small Received November 2001; accepted December 2001}

\maketitle

\section*{Abstract}

The evolution of the luminosity of nova shells in the century following
the nova outburst is studied for the lines \Hza, \Hzb, and \OzIII. About
1200 flux measurements from 96 objects have been collected from the
literature, from unpublished observations, from the HST archive, or
from new narrow-band filter imaging. For most objects, the distance and
reddening is known (or newly determined), and luminosities were calculated
from the observed fluxes. The luminosity data were combined in five
groups, according to nova light curve type (very fast, fast,
moderately fast, slow, recurrent); some objects were re-assigned to
other groups for a better fit of the luminosity data to the general trend.

For very fast, fast and moderately fast novae, the slope of the \OIII 
decline is very similar, leading to a basic `switchoff' of \OIII emission
after 11, 23 and 24 years, respectively. For the same speed classes, the
slope of the Balmer luminosity is quite similar.

In contrast to all types of fast novae, the decline in Balmer luminosity 
is more rapid in slow novae. However, the slope in \OIII is more 
gentle; slow novae still show \OIII emission after 100 years. 
Thus shells of slow novae are still hot after one century; the same
applies for the shells of the very fast nova GK Per and the recurrent nova  T
Pyx, which interact with circumstellar material.

In recurrent novae, \OIII is usually inconspicuous or absent. In
objects with giant companions, the Balmer luminosity decreases very slowly
after an outburst, which may be an effect of line blending of material
from the ejecta and the giant wind. On the other hand, objects with
dwarf companions show a very rapid decline in Balmer luminosity.

\vspace{3mm}
\noindent {\bf Keywords:} novae: shells, novae: decline, cataclysmic variables
\vspace{3mm}

\section{Introduction}

In the past years and decades, several models of nova shells have been
presented in the literature. Often they were adapted to describe the
state and evolution of specific objects, and often remarkable
agreement between model and observation was achieved.  Nevertheless it
should be kept in mind that a nova shell is a rapidly evolving object,
and its properties change significantly with time.
Furthermore, a plethora of different types of novae are
observed, which is accompanied by an amazing variety of nova shells of
various morphologies and physical properties in different stages of
temporal development.

Although studies of nova shells have been carried out since the first
bright nova of the 20th century, GK Persei in 1901, most of these
studies were carried out in a qualitative way. This approach permitted
the calculation of nebular expansion parallaxes and the morphological
study of shells. Since the shells were usually faint, and the
observations were carried out with photographic plates, hardly any
quantitative results are available. Only in the first phases of the
outburst, when the shells document themselves in the form of
emission lines, were the line fluxes estimated and derived for a few
cases, notably by Payne-Gaposchkin and collaborators.  Replacement of
the photographic plate by digital receivers has facilitated the task
of studying the evolution of nova remnants, both spectroscopically and
by means of direct imaging through narrow-band filters.  In fact,
quite a number of studies have even been carried out for extragalactic
novae, where \Hza{}-images can more easily detect the objects above the
stellar background (see, e.g. Ciardullo et al. 1987).

In this paper, we report on the results of a recent imaging survey of
nova remnants, carried out at the Kitt Peak and ESO La Silla
observatories.  We also use a hitherto unpublished survey of nova
shells carried out in 1984 at Calar Alto, and the images from the \HST
archive.  Furthermore, we have collected and homogenized the existing
quantitative record of nova shell observations.  Because the survey
attempted to cover as many objects in as many evolutionary stages
as possible, hardly any detailed information on a given object, or any
detailed modelling of shells will be given (i.e. the distribution of
line flux between various specific parts of a nova shell).  We rather
attempt to describe the ``average'' or global evolutionary track of a
nova shell, in order to derive expected values for faint shells of
ancient novae. A theoretical interpretation of the observed behavior
will be the subject of a forthcoming paper (Duerbeck \& Downes 2002).

Section 2 describes our observations and
reductions. Section 3 briefly describes the classification of novae
according to speed class, which is the base for merging our shell
luminosity data into groups. Section 4 gives the derivation of
global trends in luminosity evolution for the lines 
\Hza{}, \Hzb{} and \OIII{} in novae of
different speed classes (including, besides classical novae, recurrent
ones). Section 5 summarizes our results.

\section{Observations}

Old data of nova shell line fluxes, derived both from spectral observations or
direct images, were collected from the literature. Besides many data scattered
in the literature, the early photographic studies of Payne-Gaposchkin and 
collaborators deserve special mentioning, as well as the recent Tololo nova
survey, carried out by Williams and collaborators, and kindly put at our
disposal by him.

The new observations were obtained at the European Southern
Observatory, La Silla, Chile, and at the the Kitt Peak National
Observatory.  On 1998~March~21 -- 23, the Dutch 0.9~m telescope at ESO,
equipped with a TEK TK512CB chip ($512 \times 512$ pixels) with a
scale of $0\farcs 465$ pixel$^{-1}$ was used. On 1998~May~28 -- June~1,
observations were obtained with the KPNO 2.1~m telescope using the TEK
``T1KA'' chip ($1024 \times 1024$ pixels with a scale of $0\farcs 305$
pixel$^{-1}$), and on 1998~June~30 with the KPNO 0.9~m telescope using
the TEK ``T2KA'' chip ($2048 \times 2048$ pixels with a scale of
$0\farcs 7$ pixel$^{-1}$). A final run was carried out
at the Danish 1.54~m telescope at ESO on 2000~July~16. 
The DFOSC was used, which has a
LORAL/LESSER chip ($2052\times 2052$ pixels with a scale of $0\farcs
39$ pixel$^{-1}$).

The data were obtained with 
narrow-band filters centered
at \Ha (80 and 62 \AA~FWHM at the ESO Dutch and Danish, 36\AA~at KPNO) 
and \OIII (55 and 57\AA~at the ESO Dutch and Danish,
31\AA~at KPNO), as well as off-band and {\it UBVR} filters; see
Downes \& Duerbeck (2000) for details.  Note that the 
offband \OIII filter for the
objects observed at ESO is a Str\"omgren $y$ filter.  The data were reduced
in the standard manner.

Flux calibration of the novae were obtained via ``standard'' planetary
nebulae. NGC~6833 was used for the KPNO observations.  The
\Ha and $\rm [O~III]$ flux of NGC~6833 was determined by Tony Keyes
(private communication) based on \HST Faint Object Spectrograph
observations, and we adopted values of $3.9 \times 10^{-12}$ 
erg cm$^{-2}$ s$^{-1}$ arcsec$^{-2}$ and $9.4
\times 10^{-12}$ erg cm$^{-2}$ s$^{-1}$ arcsec$^{-2}$ 
for H$\alpha$ and $\rm [O~III]$, respectively.  For the 
``Dutch'' observations, Sp 1 was used. The
\Ha and $\rm [O~III]$ flux of Sp 1 was measured by Perinotto et al. (1994) and
the authors of the Strasbourg/ESO catalog of galactic planetary
nebulae (Acker et al. 1992) spectroscopically, and by Webster (1969),
Copetti (1990) and Shaw \& Kaler (1989) through interference filters. 
Unfortunately,
the results show some scatter, so we assume fluxes of $2.6\pm 0.5
\times 10^{-11}$ erg cm$^{-2}$ s$^{-1}$ arcsec$^{-2}$ and 
$2.8\pm 0.2 \times 10^{-11}$ erg cm$^{-2}$ s$^{-1}$ arcsec$^{-2}$ 
for H$\alpha$ and $\rm [O~III]$, respectively. For the ``Danish'' 
observations, three objects from the list of Dopita \& Hua (1997)
were used: PN 327.5+13.3, PN 327.1$-$01.8, and PN 321.3$-$16.7.

On the second night of the ESO ``Dutch'' observations, the flux of Sp 1 was
fainter by $0\fm 38$ and $0\fm 28$ in H$\alpha$ and $\rm [O~III]$, an effect
that was traced in the magnitudes of stars measured in these
filters as well. Thus, we assume that the overall transmission, and not the
central wavelength, underwent a change. Broadband transmission was
very similar in the three nights.  We thus took the nightly aperture
magnitude of the PN, corrected to airmass 1, as the ``zero-point'' of
the system, which corresponds to the fluxes given.

The targets were partly starlike, and normal {\sc Daophot} photometry was
carried out for most of the nebulae. In the $\rm [O~III]$ data,
V842~Cen and V1974~Cyg were slightly resolved, and RR~Pic and CP~Pup
were clearly resolved (Downes \& Duerbeck 2000). In these cases, the onband and
offband frames were aligned, the offband frame scaled appropriately to cancel
the majority of stars, and subtracted from the onband frame.  In most
cases, a starlike center remained (because of emission in the region
of $\rm [O~III]$), and this central emission was again fitted with a
psf-profile of the onband frame, yielding an emission-line magnitude of
the central star. The detached nebular shell emission was measured in
an annulus, where the central ``remnant'' that remained from the
subtraction and fitting attempts, was duly neglected.  More objects
were slightly resolved in \Hza, and a similar process was performed.
 
For the starlike images, 
the nova magnitudes measured from the images are a combination of the
line flux from the shell plus a continuum flux from the stellar
remnant. This continuum flux needs to be removed if we are to
study the \Ha and \OIII line fluxes themselves.  To correct for the
continuum flux, a fit was made to all objects (excluding the nova)
in the field of the form
$$
{\rm onband-magnitude} = a_{0} + a_{1} \times {\rm offband-magnitude}.
$$
These fits allowed us to estimate the flux in the continuum of the
novae, and thus to derive net line fluxes.

For the ESO observations, the use of the broad-band Str\"omgren $y$
filter as the [O~III] off-band resulted in a more involved continuum
removal.  We note that this filter has practically the same central
wavelength as a broadband Johnson $V$ filter, without, however,
including strong emission lines like the [O~III] lines.  For all field
stars, $V$ magnitudes in the standard system are available. 
The magnitude difference for the $V$- and the (uncalibrated)
$y$-observations was determined, by establishing the
relation
$$
y = V + V_0,
$$
i.e. by determining the zero-point difference $V_0$.
Furthermore, for all stars (excluding the nova) the relation 
$$
m_{\rm O~III} = y + y_0
$$
was established and $y_0$ determined. Thus the magnitude difference between the
$V$ magnitude and the [O~III]-magnitude of an object without emission lines
is
$$
m_{\rm O~III} = V + V_0 + y_0. 
$$
Field stars of all colours scatter from these linear relations by
$\pm 0\fm 01$ to $0\fm 02$, including M-stars with TiO bands. 
Thus it is likely that unusually blue stars (like nova star continua 
in their late stages) also will not deviate significantly.

Under this assumption, the presumably emission-free $y$ magnitude 
of the nova was used to calculate its hypothetical ``emission-free'' 
$V$-magnitude, and from this magnitude, its ``emission-free'' 
[O~III]-magnitude was derived. The flux
corresponding to this magnitude was subtracted from the flux
calculated from the observed [O~III]-magnitude, yielding an
``emission-only'' [O~III] flux. In most cases, the correction 
due to stellar contamination was below 10\% of the total flux.

We also evaluated direct images of nova shells, taken in August 1984 
through \Ha and \OIII filters, with a CCD camera in the Cassegrain focus
of the Calar Alto 2.2m telescope (operated by
the Centro Astronomico Hispano-Aleman, Almeria, and the Max-Planck
Institut f\"ur Astronomie, Heidelberg; observers were H.W. Duerbeck and
W.C. Seitter). Only resolved shells were considered,
the central stars (and others in the field) were reduced with the
{\sc Daophot} allstar routine within IRAF, and aperture photometry 
was carried out using MIDAS. The planetary nebula M57 (NGC 6720) served as a 
flux standard.

Furthermore, we used flux-calibrated spectroscopic observations of
several novae, taken (by us) at several telescopes in the past 25
years. Determination of the \OIII and \Ha fluxes was straightforward,
since the shells were usually unresolved.  We were also given access
to the extensive data set obtained in the Tololo Nova Survey
(cf. Williams et al. 1994), from which direct line flux measurements were
made.  

In addition to the ground-based data, we used archival 
\OIII and \Ha images obtained with the {\it Hubble Space Telescope}.  
Finally, we collected
shell fluxes from the literature, which were derived from direct 
images or spectroscopy of nova shells.

To be able to compare line strengths from different objects, we
converted the observed fluxes to luminosities; the distances and
reddenings necessary for the conversion were mostly taken from
Downes \& Duerbeck (2000); a few additional determinations or
re-determinations are discussed in the Appendix of this paper. 
The list of objects, with adopted 
distances, reddenings, and references to these data, as well as 
references for the flux measurements,
is given in Table~1. The complete set of data are given in 
Table~2, Table~3 and Table~4 for \OzIII, 
\Hza, and \Hzb, respectively.

\section{Nova groupings}

In order to merge sparse observations of many objects into a single
diagram, a reasonable grouping of objects seems desirable. Since the 
photospheric growth and shrinking is controlled by the mass loss in the course
of the outburst, and the photospheric shrinking is reflected in the light
curve, we have found it appropriate to sort the data according to 
nova speed class.

The speed classes adopted were those defined in Payne-Gaposchkin (1957): 
very fast,
fast, moderately fast, slow and very slow, depending on the average decline
rate over the first two magnitudes ($t_2$). Her classification supersedes 
previous schemes by McLaughlin (1939) (fast, average, slow, RT Ser-type) and 
Bertaud (1948)
(fast, slow, very slow), which were usually based on $t_3$ times. 
However, Payne-Gaposchkin's classification can lead to serious errors 
in some slow novae which show a pronounced maximum of somewhat more than
two magnitudes, superimposed on a plateau of almost constant brightness 
(e.g.~RR Pic). In a few cases, we have deliberately modified the
assignment to group together similar objects. Similarly, a few reassignments
were made to better fit the shell flux data of a single object to the general
trend prevailing in a neighboring speed class. Such reassignments are noted
below. 

In addition to the speed classes, we have also tried to discriminate between
the spectral classes Fe- and He/N novae according to the criteria given by 
Williams (1992); a source of this information is
Della Valle \& Livio (1998). Finally, confirmed and probable
ONeMg novae were identified, either from recent work in the literature, or
from earlier descriptions (for an early identification of such objects,
see McLaughlin (1944)).

The objects for which shell flux data were obtained are listed in 
Table 1, together with decline times $t_2,t_3$, speed class, spectral class,
and peculiarity (i.e. ONeMg group, recurrent nova, or nova with noticeable
dust formation). 

\section{Line emission as a function of time}

The nova distances and reddenings of Table 1 and the observed nova fluxes
of Tables 2, 3 and 4 were used to calculate line luminosities, which are
also given in Tables 2, 3 and 4. Depending on the distance, reddening and type
of nova, the luminosity of the stellar remnant, as well as the light gathering
and resolving power of the telescope used for the study, there is
a lower limit for the line luminosity that can still be recorded with
certainty. In general, we have assumed a luminosity of $10^{30}$ erg~s$^{-1}$  
as the lower limit that was achieved with the telescopes used in this study. 
This should be regarded as an averaged a posteriori value. To derive a 
correct estimate of the lower limit, one would not only have to take 
into account the distance, but also source confusion in the case of 
an extended shell, and the contribution of accretion disk emission 
in the case of a pointlike object.
In the following, the nova is said to have
``switched off'' emission in a specific line if its luminosity was below 
the value $10^{30}$ erg~s$^{-1}$.

\subsection{Very fast novae}

\begin{figure}
\centerline{\psfig{figure=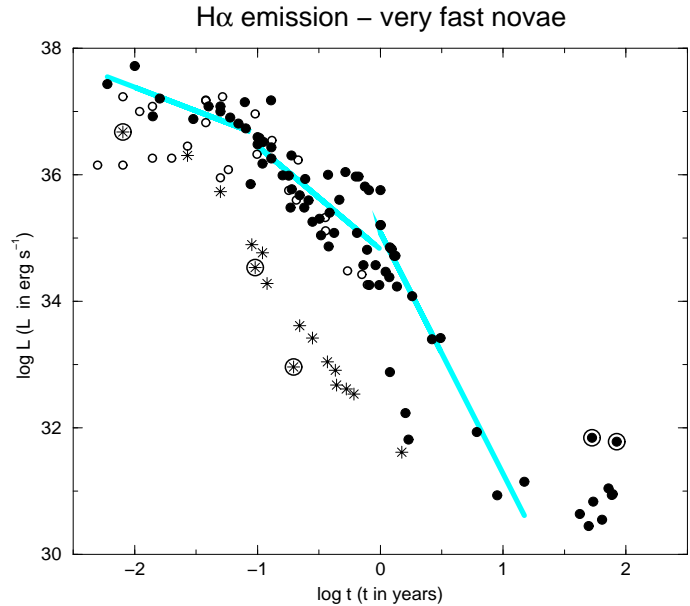,width=9cm,angle=270}}
\caption{\Ha luminosity versus time for very fast novae. Galactic
  novae are shown as filled circles, extragalactic novae as open circles. The
  unusual objects V838 Her and V4160 Sgr are shown with asterisks/encircled
  asterisks, respectively. The recent fluxes of GK Per and CP Pup are 
  shown as encircled filled circles near abscissa value 2 (GK Per is 
  the rightmost symbol).}
\end{figure}

The behavior of very fast novae is illustrated in Figs.~1 -- 3.
In these diagrams, the logarithm of the luminosity (measured in 
ergs~$\rm s^{-1}$) is plotted versus logarithm of time after maximum 
(time measured in years). Ordinary galactic novae are shown as filled
circles, and relations are determined exclusively for this galactic
dataset, unless stated otherwise. 
Extragalactic novae are shown as open circles, and are
merely displayed for comparison. Unusual novae (mentioned in the text
and the figure caption) are shown with special symbols.

\subsubsection{\rm H$\alpha$}

The \Ha evolution of very fast novae appears to be quite homogeneous,
unlike the evolution of the other spectral lines considered below. 
Several objects, however, have to be considered separately. One is
M31-C31, which is similar in evolution than the
other novae, but systematically fainter at earlier stages (see below).
The long 
series of measurements of V838 Her (indicated as asterisks in 
Figs.~1--3) shows that it is much fainter and faster
in its evolution than other very fast novae, except V4160 Sgr,
which is marked with encircled asterisks. 

Many of the famous bright novae of the 20th century (V603 Aql, GK Per
and CP Lac) are not present in this diagram, since no spectrophotometry 
was obtained longward of the photographic (blue) region.

The general evolution can be approximated by straight lines.
At times $-2.5 <\log t< -1.0$ ($<10$ days), the novae have
\pagebreak 
\begin{eqnarray*}
\log (L_{\rm H\alpha}) = &    35.89 &  -\hspace*{3mm}0.75 \log~({\rm age}),\\
                         & \hspace*{-1mm}\pm 0.38 & \hspace*{3mm}\pm 0.25
\end{eqnarray*}
or an almost constant luminosity,
$$
\log (L_{\rm H\alpha}) \approx 37.0\pm 0.4~\rm ergs~s^{-1}.
$$
At times $-1.0 < \log t < 0.0$ (10 days to 1 year)
\begin{eqnarray*}
\log (L_{\rm H\alpha}) = & 34.82 & -\hspace*{3mm}1.63 \log~({\rm age})\\
                         & \hspace*{-1mm}\pm 0.15 & \hspace*{3mm}\pm 0.25
\end{eqnarray*}
and at late times ($0.0 < \log t < 1.5$): 
\begin{eqnarray*}
\log (L_{\rm H\alpha}) = & 35.08 & -\hspace*{3mm}3.81 \log~({\rm age}). \\
                         & \hspace*{-1mm}\pm 0.11 & \hspace*{3mm}\pm 0.24
\end{eqnarray*}
The fit was restricted to $\log t < 1.5$, since at very late
times emission from the accretion disk may contaminate the measurements.
A few points were omitted, which belong to single, poorly-observed objects. 

The data points of V838 Her lie all on a straight line which can be 
approximated by
\begin{eqnarray*}
\log (L_{\rm H\alpha}) = & 31.88 & -\hspace*{3mm}2.83 \log~({\rm age}). \\
                         & \hspace*{-1mm}\pm 0.07 & \hspace*{3mm}\pm 0.09
\end{eqnarray*}

The class of `superbright novae' (V1500 Cyg, N LMC 1991) by Della Valle (1991)
is not obvious in the present diagrams.  If there is anything
peculiar, it is the {\it faintness} of M31-C31. Since
Della Valle used the M31 novae as a template (and M31-C31 appears to
fit well into this group), it appears that {\it all} very fast galactic 
novae fall into the region of the superbright novae, a fact also
clearly seen in the S-calibration of Downes \& Duerbeck (2000), as applied to
Galactic novae.  Why there is such a marked discrepancy in the
observed behaviour of very fast novae in the Galaxy and in M31
remains to be explained.

At late times (typically $\log t = 1.75$, i.e. 50 years after outburst), 
fast novae still show \Ha emission at a level of a few $ \times 10^{30}~\rm
ergs~s^{-1}$. Two objects, however, are noticeably brighter: CP Pup and GK Per.
It is obvious that the emission does not originate from the accretion disk,
since the shell is spatially resolved.

\begin{figure}
\centerline{\psfig{figure=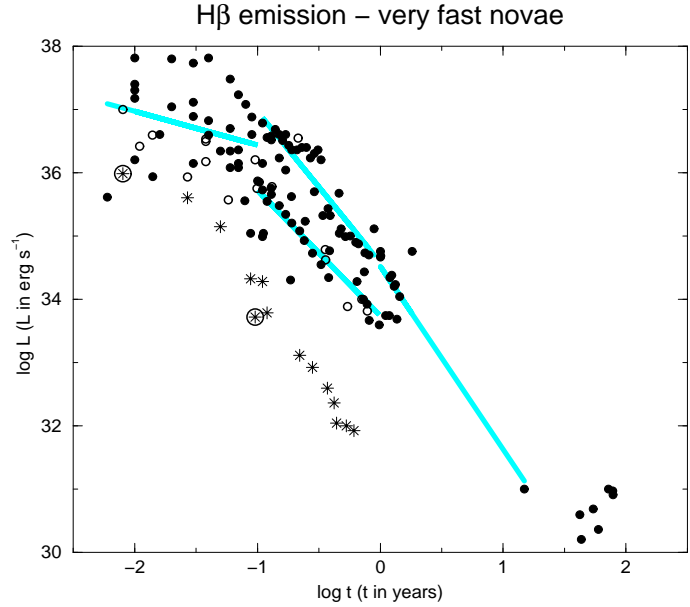,width=9cm,angle=270}}
\caption{\Hb luminosity versus time for very fast novae.  Galactic
  novae are shown as filled circles, extragalactic novae as open circles. The
  unusual objects V838 Her and V4160 Sgr are shown with asterisks/encircled
  asterisks, respectively.}
\end{figure}

\subsubsection{\rm H$\beta$}

The \Hb flux shows a large scatter (this is possibly due to the fact
that the old novae V603 Aql, GK Per, CP Lac etc. are now included). It
appears as if there are two well-defined bands, an upper and a lower
one, with only a few points in between. The extragalactic (LMC, M31)
novae seem to cluster near the lower band. As in the previous diagram,
V838 Her is notably fainter than the rest.

As in the case of \Hza, between 
$\log t= -2$ and $-1.5$ 
(i.e. 1 day and 15 days after maximum), flux levels are almost constant 
at $10^{36}$ and $10^{38}~\rm ergs~s^{-1}$ 
($<\log L> = 36.91\pm 0.71$~ergs~$\rm s^{-1}$);
afterwards, a decline sets in.
At times $-2.5 < \log t < -1.0$ 
\begin{eqnarray*}
\log (L_{\rm H\beta}) = & 35.90 & -\hspace*{3mm}0.53 \log~({\rm age}). \\
                        & \hspace*{-1mm}\pm 0.52 & \hspace*{3mm}\pm 0.35
\end{eqnarray*}
At times $-1.0 < \log t < 0.0$, a dichotomy seems indicated. The upper line is 
\begin{eqnarray*}
\log (L_{\rm H\beta}) = & 34.57 & -\hspace*{3mm}2.39 \log~({\rm age}), \\
                        & \hspace*{-1mm}\pm 0.11 & \hspace*{3mm}\pm 0.18
\end{eqnarray*}
and the lower one
\begin{eqnarray*}
\log (L_{\rm H\beta}) = & 33.74 & -\hspace*{3mm}1.98 \log~({\rm age}).\\
                        & \hspace*{-1mm}\pm 0.14 & \hspace*{3mm}\pm 0.20
\end{eqnarray*}
At late times $(0.0 < \log t < 1.5)$, the slope is only poorly documented,
and at times $\log t > 1.5$, emission from the disk may dominate. The 
slope at late times is:
\begin{eqnarray*}
\log (L_{\rm H\beta}) = & 34.52 & -\hspace*{3mm}2.89 \log~({\rm age}). \\
                        & \hspace*{-1mm}\pm 0.14 & \hspace*{3mm}\pm 0.40
\end{eqnarray*}
The time interval $(-1.0, 1.0)$ can also be fitted with one straight line:
\begin{eqnarray*}
\log (L_{\rm H\beta}) = & 34.35 & -\hspace*{3mm}1.94 \log~({\rm age}). \\
                        & \hspace*{-1mm}\pm 0.18 & \hspace*{3mm}\pm 0.27
\end{eqnarray*}
Finally, the nova V838 Her follows its own relation:
\begin{eqnarray*}
\log (L_{\rm H\beta}) = & 31.25 & -\hspace*{3mm}2.90 \log~({\rm age}). \\
                        & \hspace*{-1mm}\pm 0.08 & \hspace*{3mm}\pm 0.10
\end{eqnarray*}

The X-ray turnoff time of such objects is typically of the order one
to several hundred days according to Vablandingham et al. (2001), 
so that the break
in \Hb (and [O III]) around 1 year may have to do with the switching
off of nuclear burning on the central object.
At late stages ($t$ of the order of 50 years), objects cluster at some
$10^{30}~\rm ergs~s^{-1}$. No late \Hb observations exist for 
GK Per and CP Pup.

\begin{figure}
\centerline{\psfig{figure=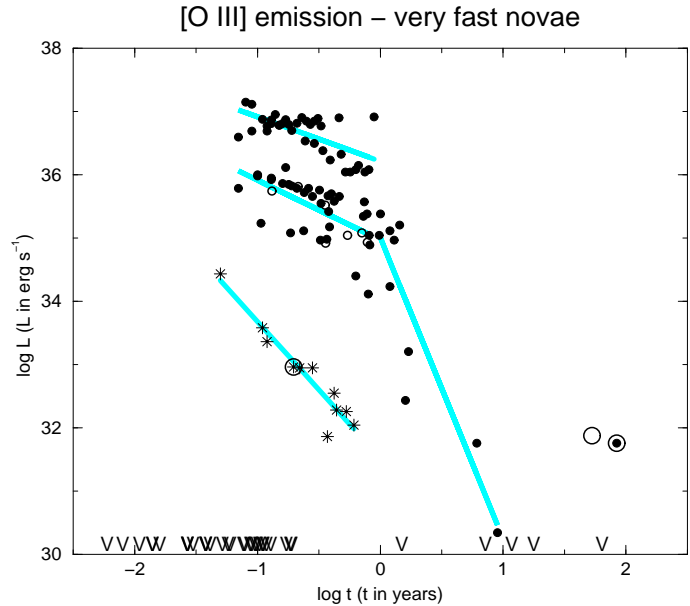,width=9cm,angle=270}}
\caption{\OIII luminosity versus time for very fast novae. Galactic
  novae are shown as filled circles, extragalactic novae as open circles. The
  unusual objects V838 Her and V4160 Sgr are shown with asterisks/encircled
  asterisks, respectively. The recent luminosity of GK Per 
  is shown as an encircled filled circle (rightmost symbol); a large open
  circle indicates that of CP Pup, which is due to N II 5005.
  Negative observations (no flux detected, upper limit about $10^{30}~\rm
  erg~s^{-1}$) are marked with the symbol ``$\vee$''.}
\end{figure}

\subsubsection{\rm [O III]}

Generally, \OIII appears around age $\log t=-1.1$ (about a month 
after optical maximum), and remains within a factor 10 constant until 
$\log t=0$ (i.e. one year after outburst). As in the case of \Hzb, 
two well-expressed bands seem to be present, with 
the extragalactic novae populating the lower one. After two years, 
the \OIII emission declines dramatically and for most 
objects is not recorded any more after 10 years. There are two noticeable
exceptions: CP Pup and GK Per, which show emission at almost $10^{32}~\rm
ergs~s^{-1}$ 50 -- 85 years after outburst. However, the luminosity 
from CP Pup is due to N II 5005, as was found spectroscopically by 
Williams (1982), while the emission in GK Per is indeed due to \OIII 
(Bode et al. 1988).

The dichotomy that appeared in the \Hb luminosities is again clearly seen.
A fit through the points with $\log t < 0$ yields 
\begin{eqnarray*}
\log (L_{5007}) = & 36.21 & - \hspace*{3mm}0.70 \log~({\rm age}) \\
                         & \hspace*{-1mm}\pm 0.09 & \hspace*{3mm}\pm 0.13
\end{eqnarray*}
for the upper line, and 
\begin{eqnarray*}
\log (L_{5007}) = & 34.97 & - \hspace*{3mm}0.94 \log~({\rm age}) \\
                         & \hspace*{-1mm}\pm 0.14 & \hspace*{3mm}\pm 0.22
\end{eqnarray*}
for the lower one.
Subsequently, both bands merge and a drop 
\begin{eqnarray*}
\log (L_{5007}) = & 35.00 & - \hspace*{3mm}4.76 \log~({\rm age}) \\
                         & \hspace*{-1mm}\pm 0.38 & \hspace*{3mm}\pm 0.88
\end{eqnarray*}
is observed. The last two data points, which belong to CP Pup and GK Per, 
were neglected in this fit. For the remaining novae, a luminosity 
of $10^{30}~\rm ergs~s^{-1}$ is reached at $\log t=1.05$, i.e.~11 
years after outburst.

The maximum phase appears to reveal the presence of two groups, which
show almost the same temporal behavior, but are clearly separated by a
factor $\sim 15$ in emission line luminosity, with an amazingly low
scatter. Among the objects in the bright group are V476 Cyg, GK Per,
V603 Aql and V977 Sco; the faint group includes CP Lac, V1500 Cyg,
V4157 Sgr, and V351 Pup.  Interestingly, both groups contain the same
mixture of Fe II/He-N novae, CO/ONeMg novae, so that there is no
obvious parameter that influences the [O III] luminosity.

A clear outlier is V838 Her, whose \OIII luminosity is about 100 times
fainter than all other objects, except V4160 Sgr. The \OIII lines 
emerge earlier than in any other object. A fit
through the points of V838 Her yields:
\begin{eqnarray*}
\log (L_{5007}) = & 31.52 & - \hspace*{3mm}2.16 \log~({\rm age}). \\
                         & \hspace*{-1mm}\pm 0.16 & \hspace*{3mm}\pm 0.24
\end{eqnarray*}
\subsubsection{Additional remarks}

The strong late emission of \Ha and \OIII in GK Per
should be pointed out. GK Per is a peculiar nova shell which seems to
interact with interstellar material or a fossil planetary nebula
(Seaquist et al. 1989). Thus
the ejected material shows shock interaction with the stationary 
circumstellar material, and remains in a hot state.

The noticeable emission in the \OIII band of CP Pup is due to N II. CP
Pup has, however, also an unusually strong \Ha luminosity. Smits (1991)
has argued for a small distance (between 525 and 850 pc) for CP Pup, which
would lower the luminosity by factors 10 and 4, respectively. The intricate
geometry (see Downes \& Duerbeck (2000) for a detailed discussion) makes these
values less likely, although they would bring the \Ha luminosity closer to
the average value found in other very fast novae of this age.

The peculiar case of V838 Her has already been mentioned. This was 
the nova with the shortest $t_2$ and $t_3$ times of the light curve decay.
V838 Her appears to have less mass ejection, and
thus a thin photosphere, which made the central object and its light
changes appear after an unusually short time (Leibowitz et al. 1992). It is 
important to note that the poorly observed nova V4160 Sgr
represents a second object of this rare class of rapidly fading neon
novae with He-spectral characteristics. The three spectra available for V4160
Sgr (Williams 2000) resemble those of V838 Her 
(Vanlandingham et al. 1996) at similar phases.
 
\begin{figure}
\centerline{\psfig{figure=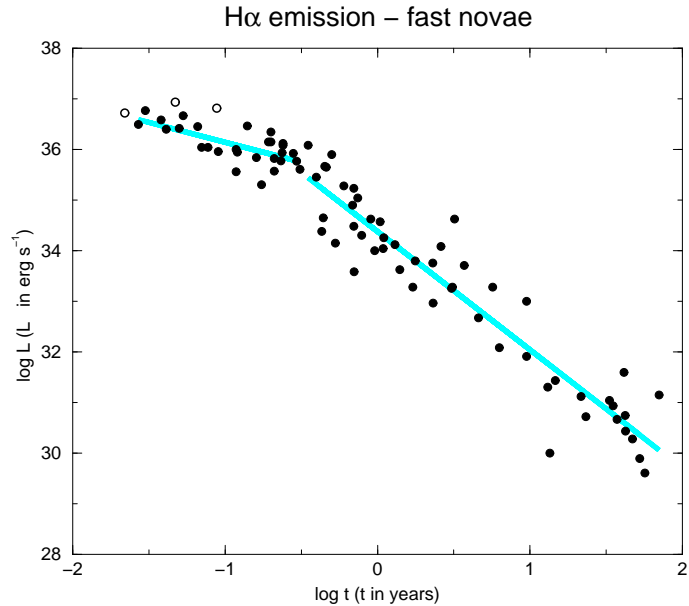,width=9cm,angle=270}}
\caption{\Ha luminosity versus time for fast novae. Galactic
  novae are shown as filled circles, extragalactic novae as open circles. }
\end{figure}

\subsection{Fast novae}

\subsubsection{\rm H$\alpha$}

The trends in the evolution of line luminosities in the class of fast novae
(Figs.~4 -- 6) differ somewhat from those of very
fast novae. The scatter between individual objects in this group is
less pronounced. There is only one extragalactic nova (N LMC 1988-1)
in the sample, which will be considered at the end.

\Ha emission is first recorded at $\log t = -1.6$ (9 days after 
maximum), and slowly declines till $\log t = -0.5$ (115 days after maximum) 
by a factor 5. After that, a steeper decline sets in.
At times $\log t < -0.5$ (115 days):
\begin{eqnarray*}
\log (L_{\rm H\alpha}) = & 35.35 & - \hspace*{3mm}0.78 \log~({\rm age}), \\
                         & \hspace*{-1mm}\pm 0.16 & \hspace*{3mm}\pm 0.16
\end{eqnarray*}
with an average value $\log (L_{\rm H\alpha})=36.08$, and at later times
\begin{eqnarray*}
\log (L_{\rm H\alpha}) = & 34.38 & - \hspace*{3mm}2.34 \log~({\rm age}).\\ 
                         & \hspace*{-1mm}\pm 0.10 & \hspace*{3mm}\pm 0.11
\end{eqnarray*}

\begin{figure}
\centerline{\psfig{figure=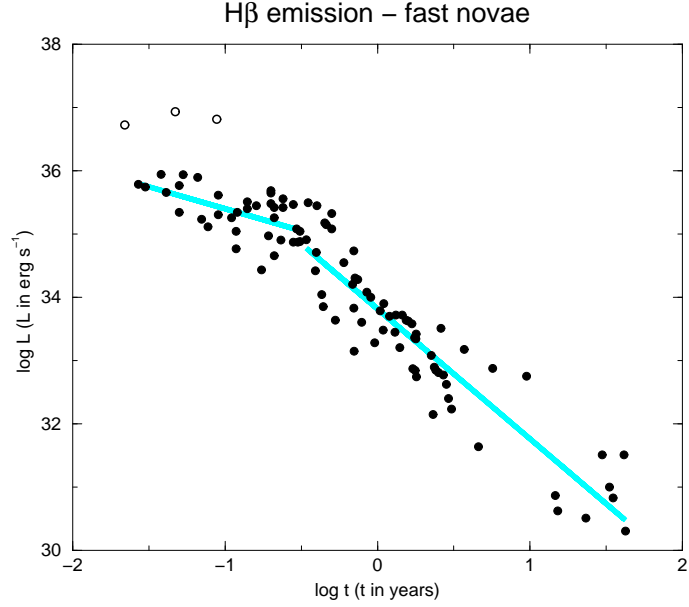,width=9cm,angle=270}}
\caption{\Hb luminosity versus time for fast novae.  Galactic
  novae are shown as filled circles, extragalactic novae as open circles.}
\end{figure}

\subsubsection{\rm H$\beta$}

\Hb emission is first recorded at $\log t = -1.7$ (7 days after 
maximum), and slowly declines till $\log t = -0.5$ (115 days after maximum) 
by a factor 3. After that, a steeper decline sets in. The extragalactic nova 
lies noticeably above the other objects. If it would be placed 
among the very fast novae (and practically all other LMC novae are very fast
novae), it would fit wery well into the corresponding diagrams. Such a case 
occurs from time to time: an object assigned to a group because of its light 
curve better fits into another group according to its line luminosity 
evolution. We will discuss the few cases in detail after the general 
discussion.
At times $\log t < -0.5$ (115 days):
\begin{eqnarray*}
\log (L_{\rm H\beta}) = & 34.70 & - \hspace*{3mm}0.69 \log~({\rm age}), \\
                         & \hspace*{-1mm}\pm 0.16 & \hspace*{3mm}\pm 0.17
\end{eqnarray*}
with an average value $\log (L_{\rm H\beta})=35.32\pm 0.38~\rm ergs~s^{-1}$, 
and at later times
\begin{eqnarray*}
\log (L_{\rm H\beta}) = & 33.81 & - \hspace*{3mm}2.05 \log~({\rm age}). \\
                         & \hspace*{-1mm}\pm 0.07 & \hspace*{3mm}\pm 0.11
\end{eqnarray*}

\begin{figure}
\centerline{\psfig{figure=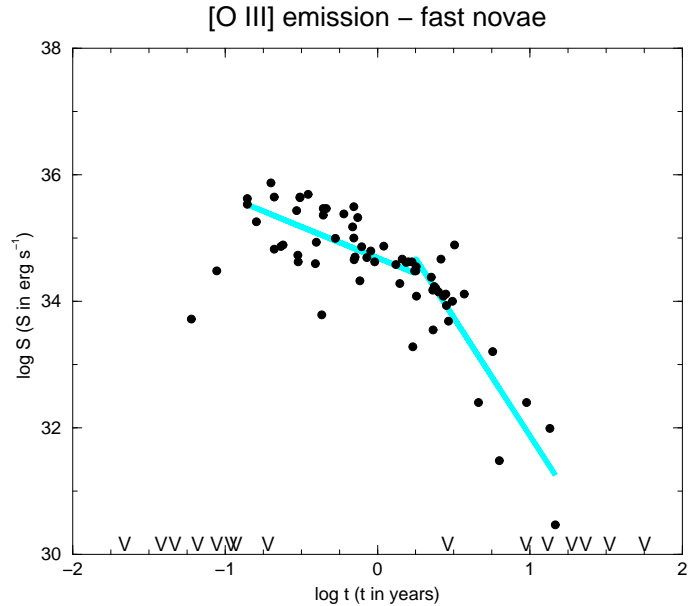,width=9cm,angle=270}}
\caption{\OIII luminosity versus time for fast novae.  Galactic
  novae are shown as filled circles, extragalactic novae as open circles.
  Negative observations (no flux detected, upper limit about $10^{30}~\rm
  erg~s^{-1}$) are marked with the symbol ``$\vee$''.}
\end{figure}

\subsubsection{\rm [O III]}

In fast novae, \OIII shows up at $\log t = -0.9$ (46 days after maximum).
It displays a continuous decline in strength (a factor of about 50) until
a breakpoint that can only be approximately determined because of the
large scatter. The breakpoint can be put at $\log t = 0$ as well as at 
$\log t = 0.5$, i.e. 1 to 3.2 years after outburst. We will assume 
$\log t = 0.25$ as the position of the breakpoint. 
After that, a steeper decline sets in.

At early stages $-1 < \log t < 0.25$, 
the average $ \log (L_{5007}) = 34.95\pm 0.53~\rm ergs~s^{-1}$,
a fit through these points with yields 
\begin{eqnarray*}
\log (L_{5007}) = & 34.68 & - \hspace*{3mm}0.98 \log~({\rm age}). \\
                         & \hspace*{-1mm}\pm 0.09 & \hspace*{3mm}\pm 0.21
\end{eqnarray*}
Subsequently, a drop
\begin{eqnarray*}
\log (L_{5007}) = & 35.60 & - \hspace*{3mm}3.73 \log~({\rm age}) \\
                         & \hspace*{-1mm}\pm 0.26 & \hspace*{3mm}\pm 0.44
\end{eqnarray*}
is observed. This means that, on the average, a luminosity of $10^{30}~\rm
ergs~s^{-1}$ is reached at $\log t=1.50$, i.e. 32 years after outburst.

\subsubsection{Additional remarks}

The X-ray turnoff time of 
ONeMg novae of the fast speed class like QU Vul and V1974 Cyg is of the 
order $1.5 - 4$ years according to Vanlandingham et al. (2001). 
This might coincide with 
the time when the \OIII emission starts to fade noticeably.

\begin{figure}
\centerline{\psfig{figure=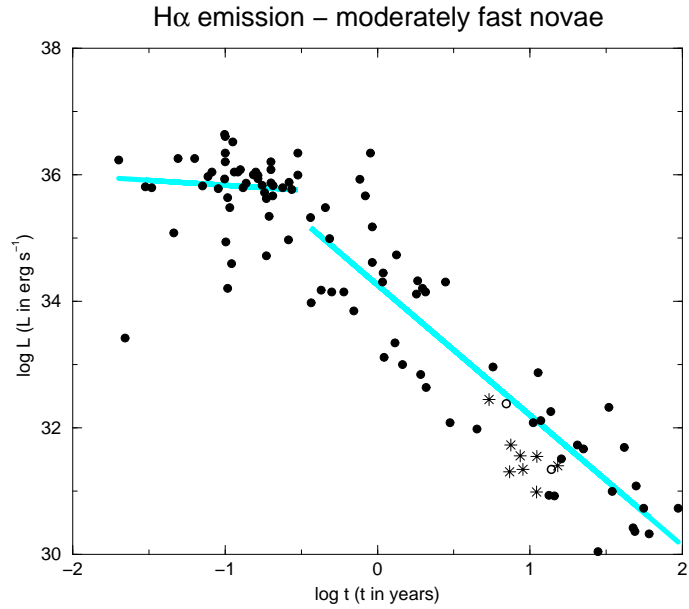,width=9cm,angle=270}}
\caption{\Ha luminosity versus time for moderately fast novae. 
Galactic  novae are shown as filled circles. The ``slow'' nova PW Vul is 
shown with open circles; it fits very well into the general trend. The
peculiar nova GQ Mus is shown with asterisks.}
\end{figure}

\subsection{Moderately fast novae}

\subsubsection{\rm H$\alpha$}

Some moderately fast novae have been caught in very early stages, 
with fairly strong or weak \Ha emission lines, but the sample is too 
small to draw any conclusions. Between $\log t = -1.5$ and $-0.5$
(12 days and 115 days), a fairly well-expressed maximum is reached.
Afterwards, a decline with a slope of $\sim -2$ sets in.
At times $\log t < -0.5$, 
\begin{eqnarray*}
\log (L_{\rm H\alpha}) = & 35.68 & - \hspace*{3mm}0.08 \log~({\rm age}), \\
                         & \hspace*{-1mm}\pm 0.27 & \hspace*{3mm}\pm 0.28
\end{eqnarray*}
with an average value $\log (L_{\rm H\alpha})=35.8\pm 0.5~\rm erg~s^{-1}$, 
and at later times
\begin{eqnarray*}
\log (L_{\rm H\alpha}) = & 34.26 & - \hspace*{3mm}2.06 \log~({\rm age}).\\ 
                         & \hspace*{-1mm}\pm 0.15 & \hspace*{3mm}\pm 0.15
\end{eqnarray*}
Figures~7 -- 9 show two objects with special symbols. Open circles denote
data for the slow nova PW Vul, which fits quite well into the group of
moderately fast novae. Asterisks denote fluxes of the peculiar nova GQ Mus.

\begin{figure}
\centerline{\psfig{figure=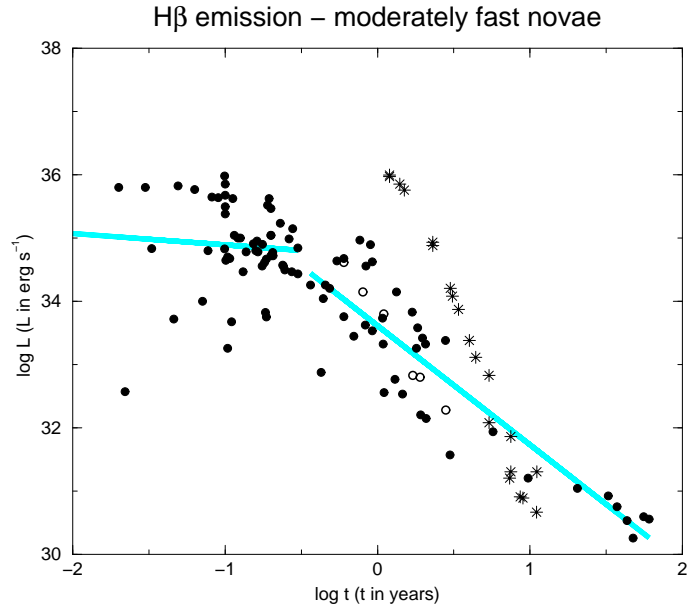,width=9cm,angle=270}}
\caption{\Hb luminosity versus time for moderately fast novae. 
 Galactic  novae are shown as filled circles. The ``slow'' nova PW Vul is 
shown with open circles; it fits very well into the general trend. The
peculiar nova GQ Mus is shown with asterisks.}
\end{figure}

\subsubsection{\rm H$\beta$}

As in the case of \Hza, in \Hb the time between $\log t = -1.5$ and $-0.5$
(12 days and 115 days) is marked by a large scatter, but no obvious
general decline is noticeable. 
Afterwards, the decline in luminosity is obvious, however also marked by large 
scatter. The slope after 4 months is $-1.78$, slightly shallower than for 
fast and very fast novae.

At times between  $\log t=-1.5$ and $\log t=-0.5$, 
\begin{eqnarray*}
\log (L_{\rm H\beta}) = & 34.71 & - \hspace*{3mm}0.18 \log~({\rm age}), \\
                         & \hspace*{-1mm}\pm 0.26 & \hspace*{3mm}\pm 0.27
\end{eqnarray*}
with an average value $\log (L_{\rm H\beta})=34.88\pm 0.68~\rm ergs~s^{-1}$, 
and at later times
\begin{eqnarray*}
\log (L_{\rm H\beta}) = & 33.61 & - \hspace*{3mm}1.88 \log~({\rm age}). \\
                         & \hspace*{-1mm}\pm 0.11 & \hspace*{3mm}\pm 0.15
\end{eqnarray*}
Again, the luminosities of the slow nova PW Vul emulate those of moderately 
fast novae, while the peculiar X-ray active nova GQ Mus shows a much more rapid
decline than other novae.

\begin{figure}
\centerline{\psfig{figure=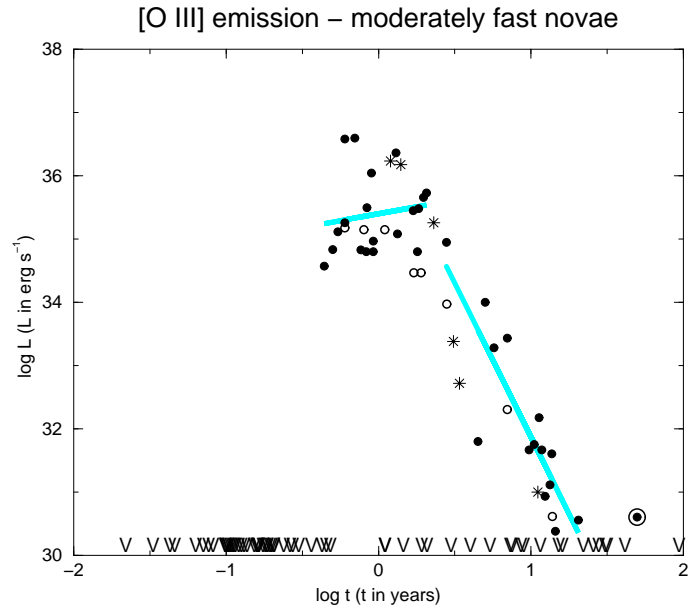,width=9cm,angle=270}}
\caption{\OIII luminosity versus time for moderately fast novae.  
Galactic  novae are shown as filled circles. The ``slow'' nova PW Vul is 
shown with open circles; it fits very well into the general trend. The
peculiar nova GQ Mus is shown with asterisks. The flux in the shell of DQ Her
at late stages is due to N II 5005 (encircled filled circle).
  Negative observations (no flux detected, upper limit about $10^{30}~\rm
  erg~s^{-1}$) are marked with the symbol ``$\vee$''.}
\end{figure}

\subsubsection{\rm [O III]}

For moderately fast novae, the \OIII data are somewhat sparse. 
\OIII shows up after $\log t = -0.4$ (145 days), and persists with 
fairly constant strength until $\log t = 0.4$ (2.5 years).
Afterwards, a quite dramatic decline sets in.

Between $-0.4 < \log t < 0.4$, the average 
$ \log (L_{5007}) = 35.39\pm 0.63~\rm erg~s^{-1}$, 
while
a fit through the points with $\log t > 0.4$, excluding the novae GQ Mus and
PW Vul and the late value of DQ Her,
yields 
\begin{eqnarray*}
\log (L_{5007}) = & 36.73 & - \hspace*{3mm}4.86 \log~({\rm age}). \\
                         & \hspace*{-1mm}\pm 0.79 & \hspace*{3mm}\pm 0.80
\end{eqnarray*}
On the average, a luminosity of $10^{30}~\rm
ergs~s^{-1}$ is reached at $\log t=1.38$, i.e. 24 years after outburst.

\subsubsection{Additional remarks}

The similarity of the evolution of the slow nova PW Vul with the majority of
moderately fast novae has already been pointed out. It is also interesting to
note that the decline in \OIII is very similar in fast and moderately fast
novae (although \OIII appears much earlier in fast novae).

An important point is the unusually bright \OIII luminosity determined 
for DQ Her 
50 years after outburst. The \OIII filter exposure has certainly picked up
emission of the line N II 5005, which was observed spectroscopically by 
Williams et al.~(1978) at the endpoints of the major axis of the shell. Indeed,
the direct image of 1984 shows a deficit of emission along the minor axis.  

\begin{figure}
\centerline{\psfig{figure=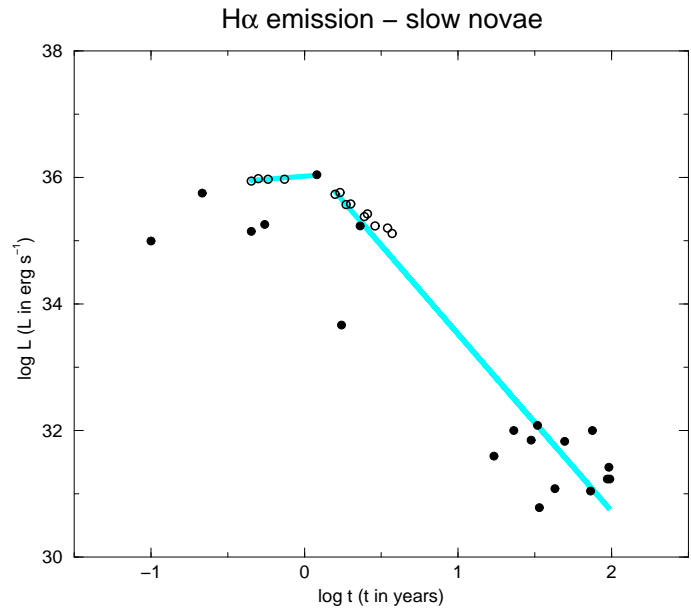,width=9cm,angle=270}}
\caption{\Ha luminosity versus time for slow novae.   Galactic
  novae are shown as filled circles, extragalactic novae as open circles.}
\end{figure}

\subsection{Slow and very slow novae}

\subsubsection{\rm H$\alpha$}

Flux data are only available for a handful of novae, making general
statements quite uncertain. The main string of data (see Figs.~10 -- 12),
is from M31-C32,
a well-monitored extragalactic very slow nova. A few data points 
from V723 Cas fall remarkably close to this object. Thus it seems that
the luminosity discrepancy between novae in M31 and the Galaxy
occurs only among very fast novae.

Note that some novae, like RR Pic, HR Del and V723 Cas have an
extended pre-maximum halt. Because of the logarithmic plot, emission
line fluxes are plotted only after maximum light. The number of
studied objects is somewhat scarce, and the behavior of slow novae is
quite varying, so these fits are not as representative as those for the other
speed classes.

The luminosity remains essentially constant between $\log t=-0.4$ and
$\log t=0.1$ ($145 - 460$ days after maximum), and declines afterwards. 

At times $-0.4 < \log t < +0.1$, 
\begin{eqnarray*}
\log (L_{\rm H\alpha}) = & 36.02 & + \hspace*{3mm}0.19 \log~({\rm age}) \\
                         & \hspace*{-1mm}\pm 0.01 & \hspace*{3mm}\pm 0.06
\end{eqnarray*}
with an average value $\log 
(L_{\rm H\alpha})=35.98\pm 0.04~\rm erg~s^{-1}$, and at later times
\begin{eqnarray*}
\log (L_{\rm H\alpha}) = & 36.33 & - \hspace*{3mm}2.80 \log~({\rm age}) \\
                         & \hspace*{-1mm}\pm 0.22 & \hspace*{3mm}\pm 0.18
\end{eqnarray*}
\Ha emission from the shell is still observed in very old remnants.

\begin{figure}
\centerline{\psfig{figure=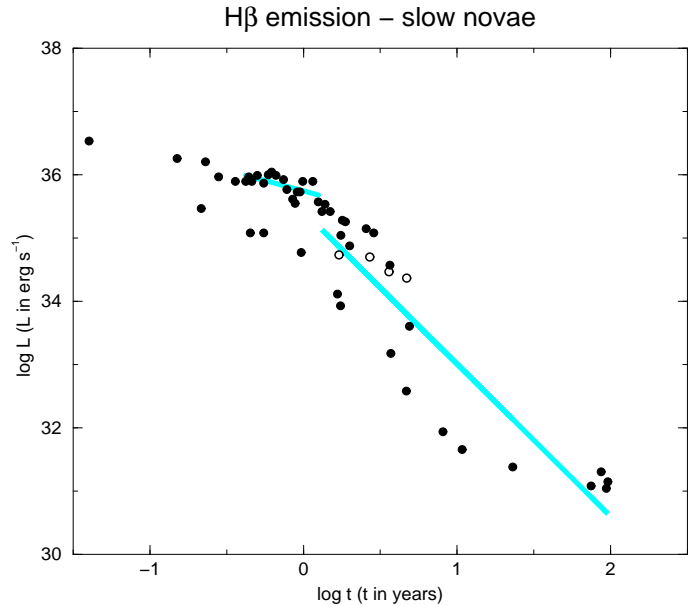,width=9cm,angle=270}}
\caption{\Hb luminosity versus time for slow novae.  Galactic
  novae are shown as filled circles, extragalactic novae as open circles. }
\end{figure}

\subsubsection{\rm H$\beta$}

Here we have a long data series of RR Pic, from the presumable
maximum onward, plus data from HR Del and V868 Cen. Between $\log t=-2$ and
$\log t=0.1$ (3 days -- 460 days after maximum),
the behavior of the \Hb flux declines very slowly 
by a factor 4. After 460 days, the slope becomes steeper.

At times $-0.4 < \log t < +0.1$, 
\begin{eqnarray*}
\log (L_{\rm H\beta }) = & 35.74 & - \hspace*{3mm}0.65 \log~({\rm age}), \\
                         & \hspace*{-1mm}\pm 0.04 & \hspace*{3mm}\pm 0.22
\end{eqnarray*}
with an average value 
$\log (L_{\rm H\beta })=35.84\pm 0.16~\rm erg~s^{-1}$, 
and at later times
\begin{eqnarray*}
\log (L_{\rm H\beta }) = & 35.42 & - \hspace*{3mm}2.41 \log~({\rm age}). \\
                         & \hspace*{-1mm}\pm 0.21 & \hspace*{3mm}\pm 0.22
\end{eqnarray*}
\Hb emission from the shell is still observed in very old remnants.

\begin{figure}
\centerline{\psfig{figure=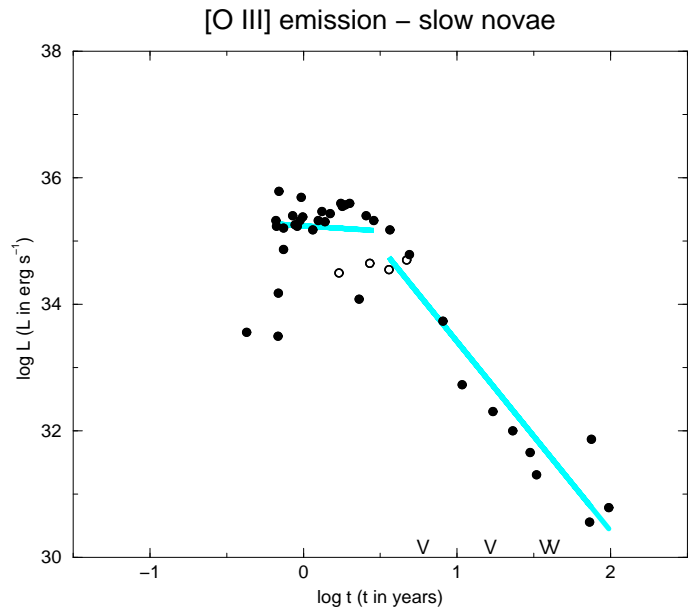,width=9cm,angle=270}}
\caption{\OIII luminosity versus time for slow novae.  Galactic
  novae are shown as filled circles, extragalactic novae as open circles. 
  Negative observations (no flux detected, upper limit about $10^{30}~\rm
  erg~s^{-1}$) are marked with the symbol ``$\vee$''.}
\end{figure}

\subsubsection{\rm [O III]}

In slow novae, \OIII emission appears at time $\log t = -0.2$, about 
230 days after maximum. RR Pic and HR Del 
show almost constant \OIII flux between the first 
appearance at 0.6 years up to $\log t = 0.5$, 3 years after maximum.

The average 
$ \log (L_{5007}) = 35.22\pm 0.43~\rm erg~s^{-1}$,
a fit through the points with $\log t < 0.5$ yields 
\begin{eqnarray*}
\log (L_{5007}) = & 35.24 & - \hspace*{3mm}0.16 \log~({\rm age}). \\
                         & \hspace*{-1mm}\pm 0.09 & \hspace*{3mm}\pm 0.43
\end{eqnarray*}
In the following century, a drop 
\begin{eqnarray*}
\log (L_{5007}) = & 36.40 & - \hspace*{3mm}3.00 \log~({\rm age}) \\
                         & \hspace*{-1mm}\pm 0.38 & \hspace*{3mm}\pm 0.29
\end{eqnarray*}
is observed.

Note that the decline from maximum is documented only by data of HR Del.
A few very old remnants, RR Pic, DO Aql and X Ser still show \OIII emission,
which is, in two cases, documented by spectroscopy (and thus confusion with N
II, as in CP Pup and DQ Her, can be ruled out). Such a behavior is at 
variance with all other types of classical novae, and indicates that 
a source of high energy photons, presumably from the nuclear burning 
in the outer layers of 
the white dwarf, is still active $\sim 100$ years after outburst. 

\subsubsection{Additional Remarks}

Most notable are the \OzIII{}-bright shells around old slow novae.

\begin{figure}
\centerline{\psfig{figure=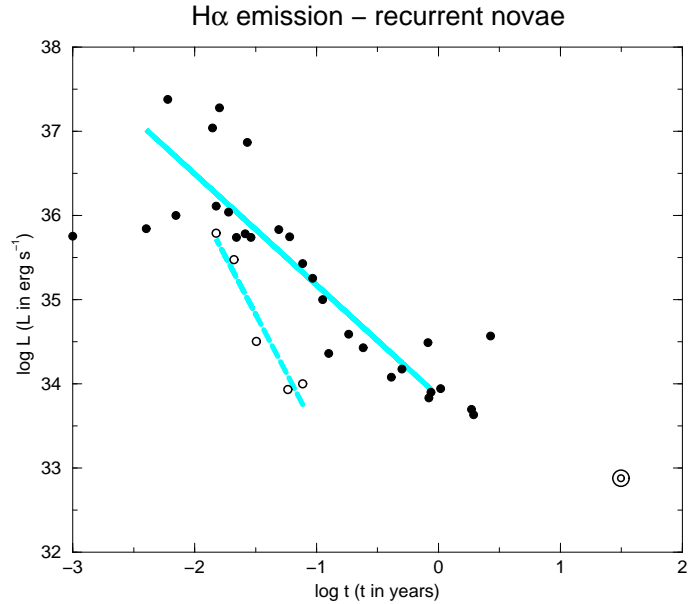,width=9cm,angle=270}}
\caption{\Ha luminosity versus time for recurrent novae. 
Galactic objects with giant companions are shown as filled circles,
galactic objects with dwarf companions as open circles. The recent flux of
the slow recurrent nova T Pyx is shown as an encircled open circle.}
\end{figure}

\subsection{Recurrent Novae}

At least three distinct groups of recurrent novae can be discriminated:
very fast novae with dwarf companions, very fast novae with giant companions,
and the slow recurrent nova T~Pyx. The most extensive data set exists for 
the second group, and it is obvious that the other groups behave differently:
the fast recurrent novae with dwarf secondaries show an extremely rapid 
decline of emission line strength in the Balmer lines, and \OIII is 
basically absent (see Figs.~13 -- 15). The slow recurrent nova T Pyx shows, 
more than thirty years
after its most recent outburst, persistently strong \Ha and \OIII emission,
likely because of shock interaction of ejecta from different outbursts.

\subsubsection{\rm H$\alpha$}

In the interval $-2.5 < \log t < 0$, recurrent novae with 
giant companions have 
\begin{eqnarray*}
\log (L_{\rm H\alpha}) = & 33.85 & - \hspace*{3mm}1.32 \log~({\rm age}).\\
                        & \hspace*{-1mm}\pm 0.22 & \hspace*{3mm}\pm 0.15
\end{eqnarray*}
Recurrent novae with dwarf companions show a more rapid decrease in flux:
\begin{eqnarray*}
\log (L_{\rm H\alpha })  = & 30.67 & - \hspace*{3mm}2.77 \log~({\rm age}). \\
                         & \hspace*{-1mm}\pm 0.67 & \hspace*{3mm}\pm 0.45
\end{eqnarray*}

\begin{figure}
\centerline{\psfig{figure=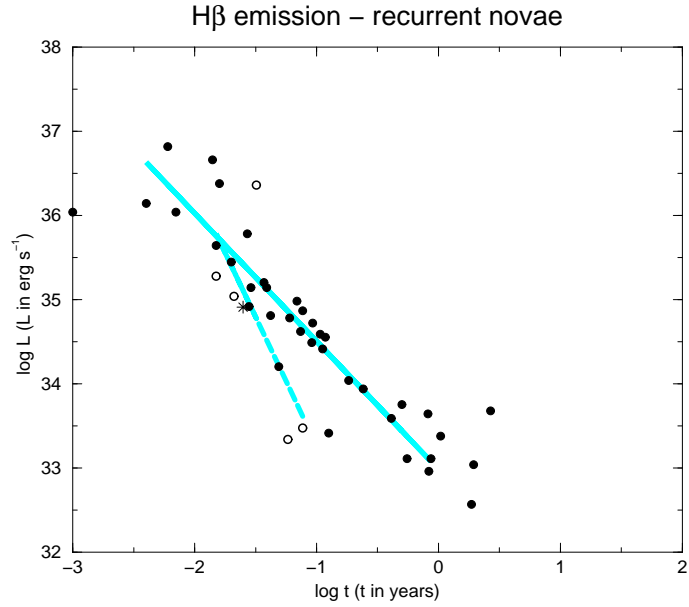,width=9cm,angle=270}}
\caption{\Hb luminosity versus time for recurrent novae.  
Galactic objects with giant companions are shown as filled circles,
galactic objects with dwarf companions as open circles; the extragalactic
recurrent nova LMC 1988-2, which also has a dwarf companion, is shown with an
asterisk.  }
\end{figure}

\subsubsection{\rm H$\beta$}

At times $-2.5 < \log t < 0$, recurrent novae with giant companions show
\begin{eqnarray*}
\log (L_{\rm H\beta }) = & 32.99 & - \hspace*{3mm}1.52 \log~({\rm age}). \\
                         & \hspace*{-1mm}\pm 0.14 & \hspace*{3mm}\pm 0.11
\end{eqnarray*}
Recurrent novae with dwarf companions show a more rapid decrease in flux:
\begin{eqnarray*}
\log (L_{\rm H\beta }) = & 30.22 & - \hspace*{3mm}3.05 \log~({\rm age}).\\ 
                         & \hspace*{-1mm}\pm 2.63 & \hspace*{3mm}\pm 1.77
\end{eqnarray*}

\begin{figure}
\centerline{\psfig{figure=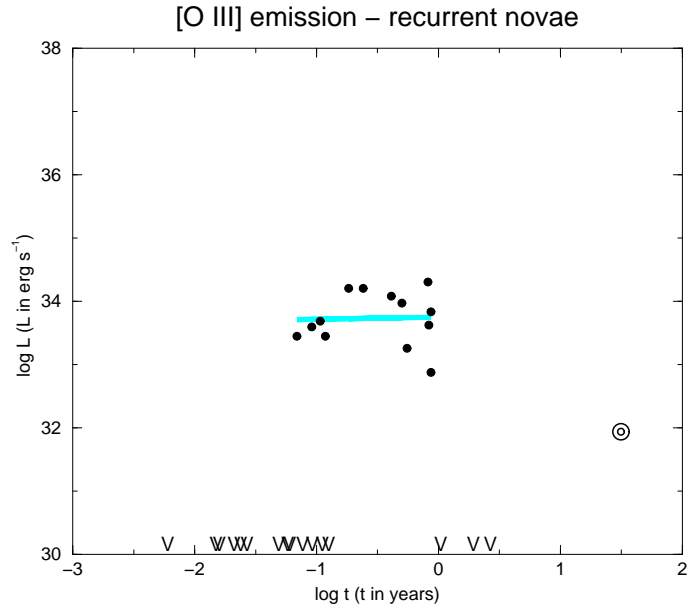,width=9cm,angle=270}}
\caption{\OIII luminosity versus time for recurrent novae.  
Galactic objects with giant companions are shown as filled circles.
Galactic objects with dwarf companions seem to show no [O III] emission;
the only exception is the slow recurrent nova T Pyx, whose recent flux 
is shown as an encircled open circle.
  Negative observations (no flux detected, upper limit about $10^{30}~\rm
  erg~s^{-1}$) are marked with the symbol ``$\vee$''.}
\end{figure}

\subsubsection{\rm [O III]}

No \OIII flux measurements in recurrent novae with dwarf companions are known;
these features are either very weak or do not appear. Early weak \OIII
emission, caused by photoionization of the giant wind by the energetic 
radiation from the early outburst, has been reported for V745 Sco.

Noticeable emission in \OIII starts to appear at $\log t = -1.2$, 
reaches a maximum at $\log t = -0.5$, and declines afterwards. 
A fit is uncertain for lack of secure data, but the average
flux at times before one year is 
$\log (L_{5007}) = 33.73\pm 0.42~\rm erg~s^{-1}$.
T Pyx is still bright in \OIII 30 years after the last outburst: 
Interaction with previous shells keeps the ejecta
hot, which appears to be untypical for other, faster
recurrent novae. Thus we refrain from merging data
of very different objects, as we have also considered 
RN with dwarf and giant companions separately.

\section{Summary}

We have collected about 1200 available line fluxes of 96 classical and
recurrent novae of various speed classes and have studied the evolution
of the luminosity in the \Hza, \Hzb, and \OIII line as a function of time
after outburst. 
In general, novae of a given speed class follow similar
patterns, so that functional relations for the average evolution of
line luminosity could be derived. 
General trends for novae of various speed classes are shown in
Figs. 15, 16 and 17.
A few novae turn out to be unusual:
GK Per and T Pyx, which interact with circumstellar material, V838 Her 
and V4160 Sgr, which have unusually small mass ejection, and GQ Mus,
an X-ray emitting classical nova.

\begin{figure}
\centerline{\psfig{figure=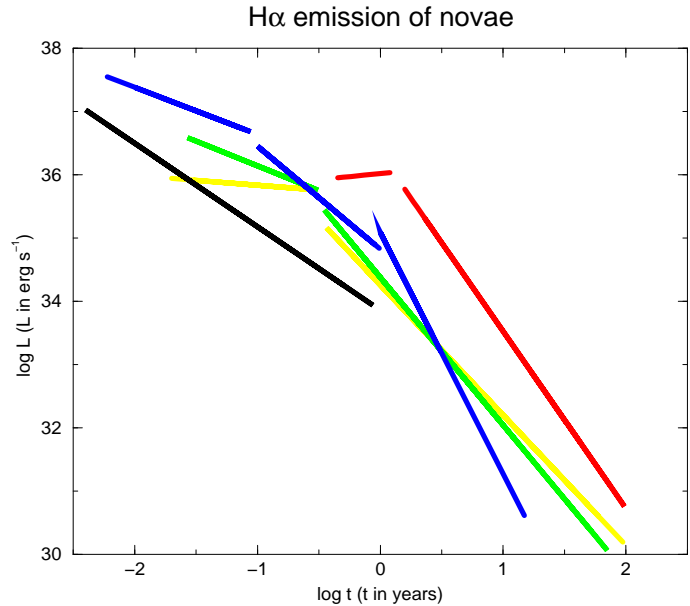,width=9cm,angle=270}}
\caption{Averaged time dependence of \Ha emission in novae of 
various speed classes: very fast (blue), fast (green), moderately fast
(yellow), slow and very slow (red), and recurrent novae with giant companions
(black). At late times, decay occurs fastest in very fast novae, somewhat
more slowly in slow novae, and slowest in fast and moderately fast novae. 
Contrary to the behaviour in classical novae, the decay of \Ha emission 
in recurrent novae sets in immediately after maximum, and continues 
without any break in slope.}
\end{figure}

\begin{figure}
\centerline{\psfig{figure=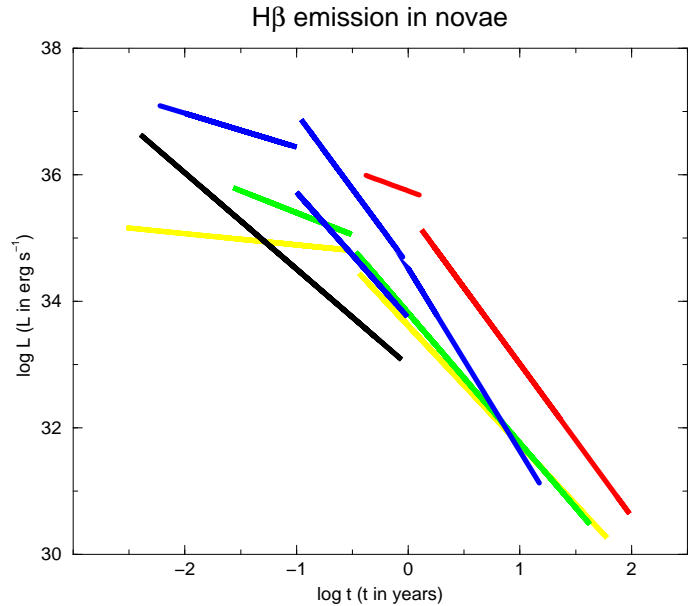,width=9cm,angle=270}}
\caption{Averaged time dependence of \Hb emission in novae of 
various speed classes: very fast (blue), fast (green), moderately fast
(yellow), slow and very slow (red), and recurrent novae with giant companions
(black). Note that the behaviour of \Hb in novae of different speed classes 
is quite similar to that shown in \Hza.  Contrary to the behaviour in 
classical novae, the decay of \Hb emission in recurrent novae sets in 
immediately after maximum, and continues without any break in slope.}
\end{figure}

\begin{figure}
\centerline{\psfig{figure=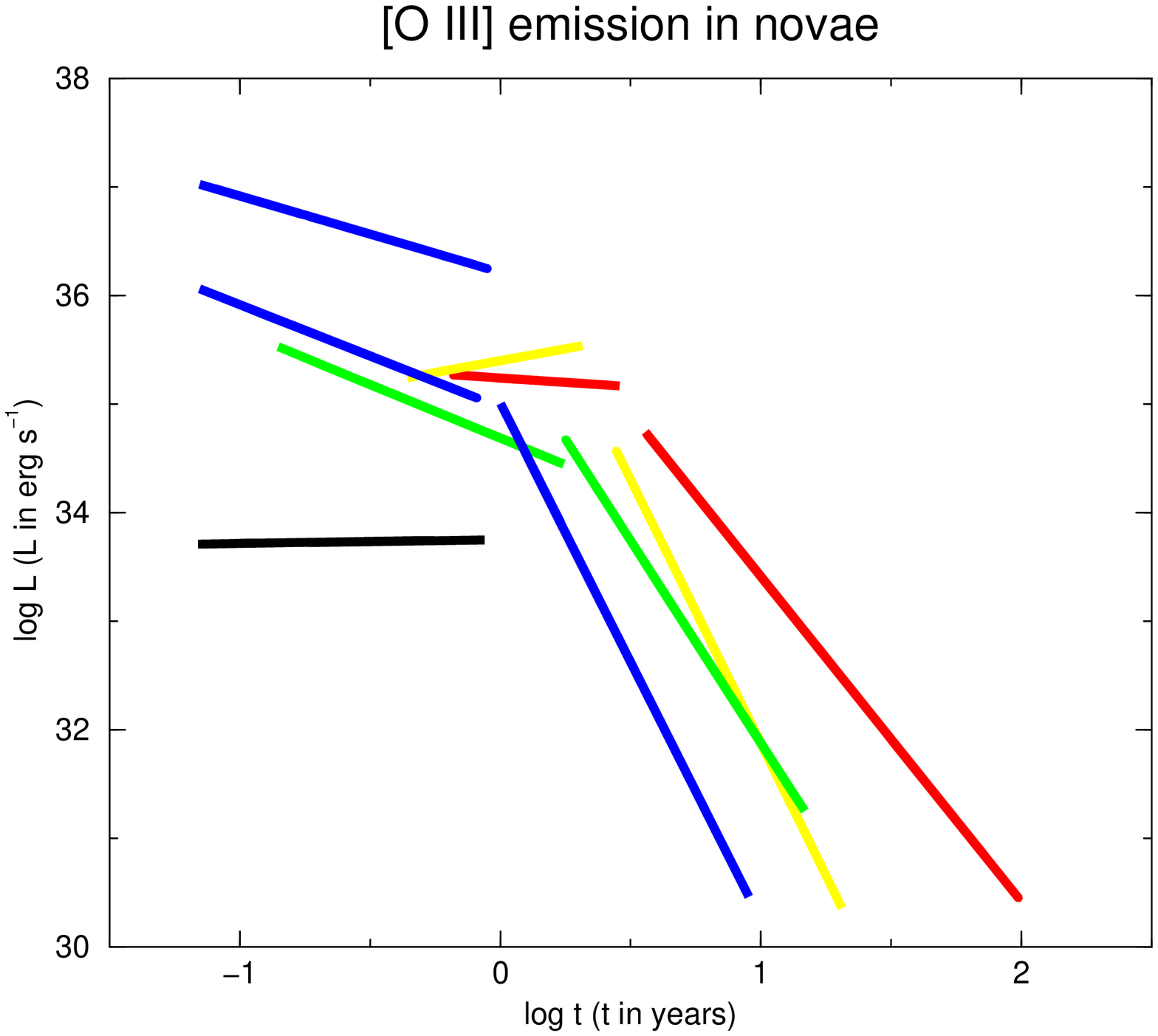,width=9cm,angle=270}}
\caption{Averaged time dependence of \OIII emission in novae of 
various speed classes: very fast (blue), fast (green), moderately fast
(yellow), slow and very slow (red), and recurrent novae with giant companions
(black).  At late times, decay occurs fastest in very fast novae, almost
as fast in moderately fast and fast novae, and slowest in slow novae, i.e.
shells of slow novae show noticeable \OIII emission also at late times.
\OIII emission in recurrent novae appears to be ``on'' for some time and 
afterwards disappears quickly. }
\end{figure}

A general discussion of the material presented here in the framework of
nova properties and shell evolution will be presented elsewhere 
(Duerbeck \& Downes 2002). The data of Tables 1 -- 4 are also included as 
ascii files in {\tt table1.dat, table2.dat, table3.dat} and {\tt table4.dat}.

\section{Appendix: Notes on some nova distances}

{\bf DO Aql} - This is a very interesting slow nova; unfortunately the
light curve has a large seasonal gap, and the true maximum may have
been missed. The reddening is low even for large distances,
$A_V=0.41$. Application of the MMRD yields a distance $d=9.5$ kpc; if
DO Aql has a luminosity that is similar to the faintest nova in the
MMRD sample, a minimum distance $d=3.6$ kpc follows. A distance $d=6.5
\pm 3.0$ is plausible.

{\bf V365 Car} - The light curve and its parameters (Liller \& Henize 1975)
indicate a slow nova for which an average absolute magnitude $M_V =
-7.09$ is assumed.  Since the light curve is photographic, a $(B-V)$
correction of 0.25 was applied. The galactic extinction program 
(Hakkila et al.~1997) yields an $A_V = 2.7$ for a distance of 4.7 kpc, which
yields the expected apparent magnitude.

{\bf V868 Cen} - The nova until now had a poor light curve
coverage. There were also discrepant reports on dust formation
(Smith et al.~1995; Harrison et al.~1998). 
The photometric observations published in the
IAU Circulars were supplemented by an extended series of visual estimates
(Jones 2001). They lead us: (1) to reclassify V868 Cen as a medium fast
nova, (2) to time the dust formation at the epoch of the IR
observations of Harrison et al.~(1998) (but certainly later than the IR
spectroscopy of Smith et al.~(1995), which however, did not develop into a
dramatic visual minimum, but a short depression, similar to those
observed in NQ Vul), and (3) to derive a consistent distance and
reddening.

{\bf V888 Cen} - Light curve data were taken from the IAU Circulars,
and $m\approx 7$, $t_3=20$ days was estimated. Using the MMRD, a
consistent distance and reddening ($d = 5100$ pc, $A_V=2.65$) was
found.

{\bf BY Cir} - Light curve data were taken from the IAU Circulars, and
$m\approx 7$, $t_2=11$ days was estimated. Using the MMRD, a
consistent distance and reddening ($d = 6000$ pc, $A_V=2.31$) was
found.

{\bf V2295 Oph} - Light curve data were taken from the IAU Circulars,
and $m\approx 9.3$, was estimated. The light curve seems to indicate a
plateau, followed by dust formation, and a light curve type C was
assumed, which is typical for novae with $M_V\sim 7.09$. A consistent
distance and reddening ($d = 10$ kpc, $A_V=1.27$) was found.

{\bf V4361 Sgr} - Light curve data were taken from the IAU Circulars,
and $m\approx 10.5$, was estimated. The nova curve is fragmentary, but
the nova appears to evolve slowly, thus $M_V = -7.09$ was assumed. A
consistent distance and reddening ($d = 4600$ pc, $A_V=4.34$) was
found.

{\bf V4633 Sgr} - Light curve data were taken from Liller \& Jones (1999), and
$m\approx 7.8$, $t_3= 52$ days was estimated. Using the MMRD, an $M_V
= -7.75$ was estimated, and a consistent distance and reddening ($d =
7300$ pc, $A_V=1.23$) was found.

{\bf V4642 Sgr} - Light curve data were taken from the IAU
Circulars. The very fragmentary light curve indicates a nova which is
not fast; maximum $m\approx 10$.  The absolute magnitude was assumed
as $M=-7$. A consistent distance and reddening ($d = 6700$ pc,
$A_V=2.72$) was found.

{\bf V1141 Sco} - Light curve data were taken from the IAU
Circulars. The light curve indicates a very fast nova with $t_2\approx
3^d, t_3\approx 7^d$, The MMRD yields an absolute magnitude of
$M=-9.8$. A consistent distance and reddening ($d = 7300$ pc,
$A_V=3.98$) was found.

{\bf V1142 Sco} - Light curve data were taken from Liller \& Jones (1999). The
light curve indicates a very fast nova with $t_2\approx 8^d,
t_3\approx 12-22^d$, maximum light was assumed to be $m_v=7$.  The
MMRD yields an absolute magnitude of $M=-9.25$. A consistent distance
and reddening ($d = 6100$ pc, $A_V=2.32$) was found.

\section*{Acknowledgments}

We thank Bob Williams (STScI) for providing the data for the Tololo
Nova Survey objects, and Albert Jones for providing his series of visual
observations of V868 Cen. We also thank the referee, A. Jorissen, for
his comments.
This research has made use of NASA's Astrophysics Data
System Bibliographic Services, as well as of the SIMBAD database,
operated at CDS, Strasbourg, France. This project was supported
by the Flemish Ministry for Foreign Policy, European Affairs, Science and
Technology.

\section*{References}

\bitem Acker, A., Marcout, J., Ochsenbein, F.,
         Stenholm, B., and Tylenda, R. 1992, Strasbourg-ESO Catalogue of 
         Galactic Planetary Nebulae (ESO, Germany)
\bitem Anderson, C.M. and Gallagher, 
        J.S. 1977, \pasp, 89, 264
\bitem Andre\"a, J.,
         Drechsel, H., and Starrfield, S. 1994, \aap, 291, 869
\bitem Anupama, G.C., Duerbeck, H.W., Prabhu, 
        T.P., and Jain, S.K. 1992, \aap, 263, 87
\bitem Anupama, G.C. and Dewangan, G.C.
        2000, \aj, 119, 1359
\bitem Anupama, G.C. and Sethi, S. 1994, 
        \mnras, 269, 105
\bitem Bahng, J.D.R.,  1972, \mnras, 158, 151
\bitem Bertaud, C. 1948, \adap, 11, 1
\bitem Bode, M.F., Duerbeck, H.W., Seitter, W.C., 
        Albinson, J.S., and Evans, A. 1988, in A Decade of UV Astronomy with 
        the IUE Satellite, ESA-SP 281, Vol. 1, 183 
\bitem Bohigas, J., Echevarr\'\i a, J., 
        Diego, F., and Sarmiento, J.A. 1989, \mnras, 238, 1395
\bitem Ciardullo, R., Ford, H.C., Neill, J.D., 
        Jacoby, G.H., and Shafter, A.W. 1987, \apj, 318, 520
\bitem Ciardullo, R., Shafter, A.W., Ford, 
        H.C., Neill, J.D., Shara, M.M. and Tomaney, A.B. 1990, \apj, 356, 472 
\bitem Ciatti, F. and Rosino, L. 1974, 
         \aaps, 16, 305
\bitem Copetti, M.V.S. 1990, \pasp, 102, 77
\bitem de Freitas Pacheco, J.A.,
         Da Costa, R.D.D., and Codina, S.J. 1989, \apj, 347, 483   
\bitem Della Valle, M. 1991, \aap, 252, L9
\bitem Della Valle, M. and Duerbeck, H.W. 1993, \aap, 275, 239 
\bitem Della Valle, M., Gilmozzi, R., Bianchini, A., and Esenoglu, H. 
         1997, \aap, 325, 1151
\bitem Della Valle, M. and Livio, M.
        1998, \apj, 506, 818
\bitem Diaz, M.P., Williams, R.E., Phillips, M.M.,
         and Hamuy, M. 1995, \mnras, 277, 959 
\bitem Dopita, M.A., and Hua, C.T. 1997, 
        \apjs, 108, 515 
\bitem Downes, R.A. and Duerbeck, H.W. 
        2000, \aj, 120, 2007
Institute, Erlangen-Nuremberg University.
\bitem Duerbeck, H.W., and Downes, R.A. 2002, in preparation
\bitem Ferland, G.J., Lambert, D.L., and
         Woodman, J.H. 1986, \apjs, 60, 375
\bitem Ferland, G.J., Williams, R.E., 
         Lambert, D.L., Slovak, M., Gondhalekar, P.M., Truran, J. W., 
         and Shields, G. A. 1984, \apj, 281, 194
\bitem Gehrz, R.D., Jones, T.J., Woodward, C.E., 
         Greenhouse, M.A., Wagner, R.M., Harrison, T.E., Hayward, T.L., and 
         Benson, J. 1992, \apj, 400, 671
\bitem Gilmozzi, R., 
         Selvelli, P., and Cassatella, A. 1998, Ultraviolet Astrophysics
         beyond the IUE final archive, ESA SP-413, eds. W. Wamsteker and
         R. Gonz\'alez Riestra, p. 415 
\bitem Gonz\'alez-Riestra, R. 1992, 
        \aap, 265, 71
\bitem Hachisu, I. and Kato, M. 2000a, 
        \apj, 540, 447  
\bitem Hachisu, I. and Kato, M. 2000b,
        \apj, 536, L93
\bitem Hachisu, I., Kato, M., Kato, T., 
        Matsumoto, K. 2000, \apj, 234, L189
\bitem Hakkila, J., Myers, J.M., Stidham, B.J., 
        and Hartmann, D.H. 1997, \aj, 114, 2043
\bitem Hang, H., Zhu, Z., and Liu, Q. 1999,
        Acta Astron. Sinica 40, 247
\bitem Harrison, T.E., Johnson, J.J., Mason, 
        P.A., and Stringfellow, G.S. 1998, in ASP Conf. 137, Wild Stars in 
        the Old West, ed. S. Howell, E. Kuulkers, and Chick Woodward 
        (San Francisco: ASP), 489
\bitem Hauschildt, P.H., Starrfield, S.,
         Shore, S.N., Gon\'zalez-Riestra, R., Sonneborn, G., and
         Allard, F. 1994, \aj, 108, 1008
\bitem Honeycutt, R.K., Robertson, J.W., and
         Turner, G.W. 1995, \apj, 446, 838
\bitem Jones, A. 2001, private communication
\bitem Kalu\.zny, J. and Chlebowski, T.
         1988, \apj, 332, 287     
\bitem Kamath, U.S., Anupama, G.C., Ashok, 
        N.M.,  and Chandrasekhar, T. 1997, \aj, 114, 2671
\bitem Kawabata, K.S., Hirata, R., Ikeda, Y.,
        Akitaya, H., Seki, M., Matsumura, M., and Okazaki, A. 2000, \apj,
        540, 429
\bitem Larsson-Leander, G. 1954, Stockholm 
        Obs. Ann. 18, No. 4
\bitem Leibowitz, E.M., Mendelson, H.,
        Mashal, E., Prialnik, D., and Seitter, W.C. 1992, \apj, 385, L49
\bitem Liller, W. and Henize, K.G. 1975,
        \apj, 200, 694
\bitem Liller, W. and Jones, A. 1999, 
        \ibvs, 4664
\bitem Liu, W., and Hu, J.Y. 2000, \apjs, 128, 387
\bitem Lynch, D.K., Rossano, G.S., Rudy, R.J., 
         and Puetter, R.C. 1995, \aj, 110, 227
\bitem Lynch, D.K., Rudy, R.J., Rossano, G.S.,
         Erwin, P., and Puetter, R.C. 1989, \aj, 98, 1682
\bitem Matheson, T.,  Filippenko, A.V., and  
         Ho, L.C. 1993, \apj, 418, L29
\bitem McLaughlin, D.B. 1939, Pop. Astr., 47, 410,
         481, 538
\bitem McLaughlin, D.B. 1944, \aj, 51, 20
\bitem McLaughlin, D.B. 1953, \apj, 117, 279   
\bitem Morisset, C. and 
         P\'equignot, D. 1996, \aap, 312, 135
\bitem Munari, U., Goranskij, V.P., Popova, A.A. 
        et al. 1996, \aap, 315, 166
\bitem Munari, U., Yudin, B.F., Kolotilov, 
         E.A., Shenavrin, V.I., Sostero, G., and Lepardo, A. 1994, 
         \aap, 284, L9
\bitem Mustel, E.R. and Boyarchuk, A.A. 
         1970, \apss, 6, 183
\bitem Payne-Gaposchkin, C. 1957, 
         The Galactic Novae (Interscience Publishers, Inc.: New York)
\bitem Payne-Gaposchkin, C. and
         Menzel, D.H. 1938, Harvard College Circ. 428
\bitem Payne-Gaposchkin, C. and
         Gaposchkin, S. 1942, Harvard College Circ. 445
\bitem P\'equignot, D., Petitjean, P.,
         Boisson, C., and Krautter, J. 1993, \aap, 271, 219
\bitem Perinotto, M., Purgathofer, A., 
         Pasquali, A., and Patriarchi, P. 1994, \aaps, 107, 481
\bitem Popper, D.M. 1940, \apj, 92, 262
\bitem Rafanelli, P. Rosino, L., and
         Radovich, M. 1995, \aap, 294, 488
\bitem Retter, A., Leibowitz, E.M., and 
        Naylor, T. 1999, \mnras, 308, 140
\bitem Ringwald, F.A. 
         Naylor, T. and Mukai, K. 1996, \mnras, 281, 192
\bitem Rosino, L., Iijima, T., Benetti, S.,
         D'Ambrosio, V., Di Paolantonio, A., and Kolotilov, E.E. 
         1992, \aap, 257, 603 
\bitem Saizar, P., Pachoulakis, I., Shore, S.N.,
         Starrfield, S., Williams, R.E., Rothschild, E., and Sonneborn, G., 
         1996, \mnras, 279, 280
\bitem Saizar, P., Starrfield, S., Ferland, G.J.,
         Wagner, R.M., Truran, J.W., Kenyon, S. J., Sparks, W. M., 
         Williams, R. E., and Stryker, L. L. 1991, \apj, 367, 310
\bitem Saizar, P., Starrfield, S., Ferland, G.J.,
         Wagner, R.M., Truran, J.W., Kenyon, S.J., Sparks, W.M., 
         Williams, R.E., and Stryker, L.L. 1992, \apj, 398, 651
\bitem Schwarz, G.J., Hauschildt, P.H.,
         Starrfield, S., Baron, E., Allard, F., Shore, S.N., and
         Sonneborn, G. 1997, \mnras, 284, 669 
\bitem Scott, A.D., Evans, A., and
         Rawlings, J.J.M. 1994, \mnras, 269, L21
\bitem Scott, A.D., Duerbeck, H.W., Evans, A., Chen, A.-L., de Martino, D., 
Hjellming, R., Krautter, J., Laney, D., Parker, Q.A., Rawlings, J.M.C., 
and Van Winckel, H. 1995, \aap, 296, 439
\bitem Seaquist, E.R., Bode, M.F., Frail, D.A.,
         Roberts, J.A., Evans, A., and Albinson, J.S. 1989, \apj, 344, 805 
\bitem Sekiguchi, K., Whitelock, P.A., Feast, M.W., Barrett, P.E.,
Caldwell, J.A.R., Carter, B.S., Catchpole, R.M., Laing, J.D., Laney, C.D.,
Marang, F., and van Wyk, F. 1990a, \mnras, 246, 78
\bitem Sekiguchi, K., Stobie, R.S., 
        Buckley, D.A.H., and Caldwell, J.A.R. 1990b, \mnras, 245, 28p
\bitem Shafter, A.W. 1997, \apj, 487, 226
\bitem Shaw, R.A. and Kaler, J.B. 1989, \apjs, 
         69, 495 
\bitem Shore, S.N. Sonneborn, George;
 Starrfield, Sumner G.; Hamuy, M.;
 Williams, R. E.; Cassatella, A.; Drechsel, H.et al. 1991, \apj, 370, 193
\bitem Smits, D.P. 1991, \mnras, 248, 217
\bitem Smith, C.H., Aitken, D.K., Roche, P.F., 
        and Wright, C.M. 1995, \mnras, 277, 259
\bitem Snijders, M.A.J., Batt, T.J., 
        Seaton, M.J., Blades, J.C., and Morton, D.C. 1984, \mnras, 211, 7p
\bitem Snijders, M.A.J., Batt, T.J.,
        Roche, P.F., Seaton, M.J., Morton, D.C., Spoelstra, T.A.T., 
        and Blades, J.C. 1987, \mnras, 228, 329
\bitem Somers, M.W. and Naylor, T. 1999, 
         \aap, 352, 563
\bitem Szkody, P. and Howell, S.B.  1992, 
         \apjs, 78, 537      
\bitem Thorstensen, J.R. and Taylor, 
        C.J. 2000, \mnras, 312, 629
\bitem Tolbert, C.R., Pecker, J.C., and 
        Pottasch, S.R. 1967, \bain, 19, 17
\bitem Tomaney, A.B. and Shafter, R.W. 1993,
        \apj, 411, 640
\bitem Tylenda, R. 1977, Acta Astr., 27, 389
\bitem Vanlandingham, K.M, Starrfield, S., 
         Wagner, R.M., Shore, S.N., and Sonneborn, G. 1996, \mnras, 282, 563
\bitem Vanlandingham, K.M, Schwarz, G.J., 
        Shore, S.N., and Starrfield, S. 2001, \aj, 121, 1126
\bitem Webster, B.L. 1969, \mnras, 143, 79
\bitem Weiler, E.J. and Bahng, J.D.R. 
         1976, \mnras, 174, 563
\bitem Whipple, F.H.
         and Payne-Gaposchkin, C. 1937, Harvard College Circ. 414
\bitem Whipple, F.H.
         and Payne-Gaposchkin, C. 1939, Harvard College Circ. 433
\bitem Whitney, B.A. and Clayton, G.C.
         1989, \aj, 98, 297
\bitem Williams, R.E. 1982, \apj, 261, 170
\bitem Williams, R.E. 1992, \aj, 104, 725
\bitem Williams, R.E. 1994, \apj, 426, 279
\bitem Williams, R.E. 2000, private communication
\bitem Williams, R.E., Woolf, N.J., Hege, E.K.,
         Moore, R.L., and Kopriva, D.A. 1978, \apj, 224, 171
\bitem Williams, R.E., Phillips, M.M., and
         Hamuy, M. 1994, \apjs, 90, 297
\bitem Zhen-xi, Z. and Heng-rong, H. 2000
          Chin. \aap, 24, 71 
\bitem Zwitter, T. and Munari, U. 1995, 
      \aaps, 114, 575 
\bitem Zwitter, T. and Munari, U. 1996, 
         \aaps, 117, 449

\newpage

\begin{table}

Table 1. Nova Characteristics
\vspace*{1mm}

\begin{tabular}{lcccccrrlll}\hline
\footnotesize
Object & t$_{2}$ & t$_{3}$ & Speed & Spectral & Pecul- & Dist. &
Ext.  & Dist. & Ext.  & Flux data \\
       & days  & days  & Class & Class    & iarity & pc\phantom{d\,} &
$E_{B-V}$ & Ref. & Ref. & References \\ \hline
OS And~~~ &  10 & 21 &  F &     &      &  5100 & 0.25 & (1) & (1) & (1,2) \\
DO    Aql & 430 &900 &  S &     &      &  6500 & 0.13 & (3) & (3) & (4)   \\
V603  Aql &   4 &  9 & VF & hy  &      &   330 & 0.07 & (5) & (5) & (2,4,6,7)\\
V1229 Aql &  20 & 38 & MF & Fe  &      &  2100 & 0.50 & (5) & (5) & (8) \\
V1370 Aql &   9 & 13 &  F &     &      &  5000 & 0.6  & (90 & (9) & (2,10) \\
V1419 Aql &  19 & 32 &  F &     &      &  2800 & 0.78 & (3) & (11,12) & (13)\\
V1425 Aql &  11 & 23 &  F & Fe  &      &  2700 & 0.76 & (14) & (14) & (14) \\
T     Aur &  45 & 50 & MF & Fe  & dust &   960 & 0.21 & (5) & (5) & (3,6,15)\\
V365  Car & 330 &530 &  S &     &      &  4700 & 0.85 & (3) & (3) & (4)  \\
V705  Cas &  65 &100 & MF &     &      &  3200 & 0.50 & (16) & (16) & (13,17)\\
V723  Cas &  19 &180 &  S &     &      &  3000 & 0.45 & (18) & (18) & (19)  \\
V812  Cen &     &    &    &     &      & 10500 & 0.44 & (20) & (20) & (4)   \\
V842  Cen &  35 & 48 & MF & Fe  & dust &  1150 & 0.55 & (5) & (5) & (4,21,22)\\
V868  Cen &  55 &  ? & MF & Fe  &      &  1900 & 1.75 & (3) & (23) & (4,13) \\
V888  Cen &  11 & 20 &  F &     &      &  5000 & 0.84 & (3) & (3) & (4,24) \\
IV    Cep &  16 & 36 &  F &     &      &  5300 & 0.63 & (20) & (20) & (6,25) \\
BY    Cir &  11 & 50 &  F &     &      &  6000 & 0.73 & (3) & (3) & (4) \\
CP    Cru &   4 &    & VF &     &      &  3200 & 0.63 & (5) & (5) & (4)  \\
Q     Cyg &   5 & 11 & VF &     &      &  1700 & 0.25 & (20) & (20) &  (6) \\
V450  Cyg &  95/88? &108 & MF & Fe  & &  3500 & 0.41 & (5) & (5) & (2)  \\
V476  Cyg &   6 & 15 & VF & Fe  &      &  1620 & 0.26 & (5) & (5) & (2,7,15,26)\\
V1500 Cyg &   2 &  4 & VF & hy  &      &  1500 & 0.50 & (5) & (5) & (2,27,28)  \\
V1819 Cyg &  37 & 89 & MF & Fe  &      &  7400 & 0.35 & (5) & (5) & (4,29)  \\
V1974 Cyg &  17 & 37 &  F & Fe  & ONeMg&  1800 & 0.35 & (5) & (5) & (4,30)  \\
HR    Del & 172 &230 & VS & Fe  &      &   760 & 0.15 & (5) & (5) & (2,3,4,6,31,32)\\
DN    Gem &  16 & 37 &  F &     &      &  1600 & 0.10 & (33) & (20) & (26)  \\ 
DQ    Her &  39 & 86 & MF & Fe  & dust &   480 & 0.1  & (5) & (5) & (2,3,6,7,15,34,35,36) \\
V446  Her &   7 & 12 &  F & He  &      &  1350 & 0.36 & (5) & (5) & (6,17,37,38) \\
V533  Her &  22 & 46 &  F & Fe  &      &  1250 & 0.0  & (5) & (5) & (6,38) \\
V827  Her &  25 & 60 & MF &     &      &  8500 & 0.16 & (20) & (20) & (4) \\
V838  Her &   2 &  5 & VF &hy/He?&(neon)& 3000 & 0.49 & (39) & (39) & (13,39,40) \\
CP    Lac &   5 & 10 & VF & hy  &      &  1350 & 0.20 & (5) & (5) & (2,6,41) \\
DI    Lac &  20 & 43 & MF &     &      &  2600 & 0.16 & (20) & (20) & (26) \\
DK    Lac &  11 & 24 &  F & Fe  &      &  3900 & 0.44 & (5) & (5) & (2,6,42) \\
HY    Lup &    &$>25$& MF &     &      &  1800 & 0.22 & (5) & (5) & (4) \\
HR    Lyr &  45 & 74 &  S &     &      &  4200 & 0.16 & (20) & (20) & (15) \\
BT    Mon &   ? &  ? &  ? &     &      &  1800 & 0.15 & (5) & (5) & (6,42) \\
GQ    Mus &   5 & 45 &F-MF&     &      &  4800 & 0.42 & (43) & (43) & (4,23,43,44,45) \\
V841  Oph &  50 &145 & MF &     &      &   300 & 0.32 & (20) & (20) & (6)  \\
V972  Oph &     &176 &  S &     &      &  2200 & 0.63 & (20) & (20) & (2,4,46) \\
V2104 Oph &     &    &    &     &      &  3300 & 0.06 & (20) & (20) & (4) \\
V2214 Oph &  56 & 92 & MF & Fe  &(neon)&  1600 & 0.72 & (47) & (23,47) & (4,13)  \\
V2264 Oph &  19 & 45 &  F & Fe  &(neon)&  6000 & 0.50 & (47) & (23) & (13) \\
V2295 Oph &  13 & 21 &  F &     &      & 10000 & 0.40 & (3) & (3) & (13) \\     
GK    Per &   7 & 13 & VF & He  & neon!&   455 & 0.3 & (5) & (5) & (3,6,7,15)   \\
V400  Per &  20 &    &  F &     &      & 13100 & 0.06 & (20) & (20) & (48) \\\hline
\end{tabular}
\end{table}
\clearpage

\begin{table}

Table 1. Nova Characteristics (continued)
\vspace*{1mm}

\begin{tabular}{lcccccrrlll}\hline
\footnotesize
Object & t$_{2}$ & t$_{3}$ & Speed & Spectral & Pecul- & Dist. &
Ext.  & Dist. & Ext.  & Flux data \\
       & days  & days  & Class & Class    & iarity & pc\phantom{d\,} &
$E_{B-V}$ & Ref. & Ref. & References \\ \hline
RR    Pic &  20 &127 &  S & Fe  &      &   580 & 0.05 & (5) & (5) & (4,15,49)  \\
CP    Pup &   6 &  8 & VF & He  & neon?&  1700 & 0.25 & (5) & (5) & (4,15)  \\
HS    Pup &     & 65 & MF &     &      &  5400 & 0.35 & (20) & (20) & (50) \\
HZ    Pup &     & 70 & MF &     &      &  5800 & 0.25 & (20) & (20) & (50) \\
V351  Pup &  10 & 26 & VF & Fe  &ONeMg &  2700 & 0.72 & (5) & (5) & (4,13,51)   \\
V630  Sgr &   4 & 11 & VF &     & neon!&   600 & 0.51 &  (20) & (20) & (15)\\
V3888 Sgr &   ? &  ? &  ? &     &      &  2500 & 1.02 & (5) & (5) & (4) \\
V4157 Sgr &  10 & 22 & VF & Fe  &      &  7000 & 0.94 & (3) & (23) & (13) \\
V4160 Sgr &   6 & 14 & VF & He  &(neon)&  5200 & 0.29 & (20) & (20) & (13) \\
V4169 Sgr &  22 & 43 &  F &     &      &  6500 & 0.36 & (52) & (52) & (13) \\
V4171 Sgr &  17 & 31 &  F &     &      &  4300 & 0.60 & (3) & (23) & (13) \\
V4361 Sgr &$>30$&    &MF/S&     &      &  4500 & 1.38 & (3) & (3) & (4) \\
V4444 Sgr & 3.5 &    & VF &     &      &  4700 & 0.75 & (53) & (53) & (4) \\
V4633 Sgr &  16 & 35 &  F & Fe  &      &  7300 & 0.39 & (3) & (3) & (4) \\
V4642 Sgr &   ? &    & MF?& Fe  &      &  7200 & 0.87 & (3) & (3) & (4) \\
V960  Sco &     &    &    &     &      &       &      &     &     & (4) \\
V977  Sco &   3 &  8 & VF & Fe  &      & 10000 & 1.48 & (3) & (3) & (13) \\
V992  Sco &   ? &    & MF &     &      &  1600 & 0.72 & (3) & (3) & (13) \\
V1141 Sco &   3 &  7 & VF &     &      &  7200 & 1.26 & (3) & (3) & (4)  \\
V1142 Sco &   8 & 12 & VF &     &      &  6000 & 0.74 & (3) & (3) & (4)  \\
FV    Sct &     & ?  &    &     &      &       &      &     &     & (6)  \\
V368  Sct &  40 & 85 & MF &     &      &  3900 & 0.73 & (20) & (20) & (54) \\
V373  Sct &     &    & MF &     &      &  4600 & 0.32 & (20) & (20) & (6) \\
V443  Sct &  19 & 39 &  F & Fe  &      &  8000 & 0.27 & (55) & (55) & (13,55)  \\
V444  Sct &   6 & 10 & VF & Fe  &neon? &  8000 & 1.66 & (3) & (3) & (13) \\
X     Ser & 370 &555 & VS &     &      &  3000 & 0.25 & (38) & (38) & (4,6,38)  \\
CT    Ser &   ? &  ? & F? &     &      &  1400 & 0.01 & (5) & (5) & (2,4,6,56) \\
FH    Ser &  42 & 59 & MF & Fe  & dust &   920 & 0.64 & (5) & (5) & (2,6,57) \\
LW    Ser &  40 & 50 & MF &     &      &  7900 & 0.32 & (20) & (20) & (4) \\
XX    Tau &  24 & 42 &  F & Fe  & dust &  3500 & 0.40 & (5) & (5) &  \\
RW    UMi &  48 & 88 & MF & Fe? &      &  5000 & 0.09 & (5) & (5) & (2,4) \\
LV    Vul &  21 & 43 &  F & Fe  &      &   920 & 0.56 & (5) & (5) & (2) \\
NQ    Vul &  23 & 53 &  F & Fe  &      &  1160 & 0.67 & (5) & (5) & (2,4,6) \\
PW    Vul &  82 &126 &  S & Fe  &      &  1800 & 0.55 & (5) & (5) & (2,4,58) \\
QU    Vul &  22 & 49 &  F & Fe  &ONeMg &  1750 & 0.60 & (5) & (5) & (4,57,59,60) \\
QV    Vul &  50 & 53 & MF & Fe  &      &  2700 & 0.40 & (5) & (5) & (4,61,62) \\
\multicolumn{9}{l}{recurrent novae}\\
V394  CrA & 3.3 &  8 & VF & He  &recur &  4200 & 1.10 & (63) & (63) & (13) \\
RS    Oph &   4 & 14 & VF &     &recur &   570 & 0.73 & (64) & (64) & (65,66)\\    
T     Pyx &  62 & 88 & MF &     &recur &  4240 & 0.24 & (67) & (67) & (4) \\
V3890 Sgr &   9 & 18 & VF & He  &recur &  3600 & 1.1 & (68) & (68,69) & (13,69) \\
U     Sco & 2.0 &  4 & VF &     &recur &  7000 & 0.56 & (70) & (70) & (71) \\
V745  Sco & 6.6 & 10 & VF & He  &recur &  8000 & 1.13 & (72) & (23) & (13)\\ \hline
\end{tabular}
\end{table}
\clearpage

\begin{table}

Table 1. Nova Characteristics (continued)
\vspace*{1mm}

\begin{tabular}{lcccccrrlll}\hline
\footnotesize
Object & t$_{2}$ & t$_{3}$ & Speed & Spectral & Pecul- & Dist. &
Ext.  & Dist. & Ext.  & Flux data \\
       & days  & days  & Class & Class    & iarity & pc\phantom{d\,} &
$E(B-V)$ & Ref. & Ref. & References \\ \hline
\multicolumn{9}{l}{extragalactic novae}\\
LMC  88-1 &  20 & 43 &  F & Fe  &      & 55 kpc & 0.25 & & (23) & (13) \\
LMC  88-2 &   5 &    & VF & Fe  &      & 55 kpc & 0.25 & &      & (13) \\
LMC  90-1 &   8 &    & VF & ?   &ONeMg & 55 kpc & 0.25 & &      & (13) \\
LMC  91   &   5 &  8 & VF & Fe  &      & 55 kpc & 0.25 & & (23) & (13) \\
LMC  92   &   9 & 14 & VF &     &      & 55 kpc & 0.25 & & (23) & (13) \\
M31 C-31  &     &    & S  &     &      & 770 kpc & 0.11 & & & (73,74,75)\\
M31 C-32  &   8 & 12 & VF &     &      & 770 kpc & 0.11 & & & (73,74)\\
\multicolumn{9}{l}{extragalactic recurrent novae}\\
LMC  90-2 &   3 &  5 & VF & He  &recur & 55 kpc & 0.11 & & (76,77) & (13) \\ \hline
\end{tabular}
\end{table}

References to Table 1.\\
(1) Schwarz et al. (1997)\\
(2) Ringwald, Naylor \& Mukai (1996)\\
(3) this work (new determination)\\
(4) this work (imaging)\\
(5) Downes \& Duerbeck (2000)\\
(6) this work (spectroscopy)\\
(7) Payne-Gaposchkin \& Gaposchkin (1942)\\  
(8) Della Valle \& Duerbeck (1993)\\
(9) Snijders et al. (1984)\\
(10) Snijders et al. (1987)\\
(11) Lynch et al. (1995)\\ 
(12) Munari et al. (1994)\\
(13) Williams (2000)\\
(14) Kamath et al. (1997)\\
(15) Payne-Gaposchkin (1957)\\
(16) Hauschild et al. (1994)\\
(17) Liu \& Hu (2000)\\
(18) Munari et al. (1996)\\
(19) Hang, Zhu \& Liu (1999)\\
(20) Shafter (1997)\\
(21) Andre\"a et al. (1994)\\
(22) de Freitas Pacheco et al. (1989)\\
(23) Williams (1994)\\
(24) Zhen-xi \& Heng-rong (2000)\\
(25) Bahng (1972)\\
(26) McLaughlin (1953)\\
(27) Kalu\.zny \& Chlebowski (1988)\\
(28) Ferland et al. (1986)\\
(29) Whitney \& Clayton (1989)\\
(30) Rafanelli et al. (1995)\\
(31) Tylenda (1977)\\
(32) Anderson \& Gallagher (1977)\\
(33) Retter et al. (1999)\\
(34) Whipple \& Payne-Gaposchkin (1937, 1939)\\
(35) Ferland et al. (1984)\\
(36) Mustel \& Boyarchuk (1970)\\

\clearpage

References to Table 1 (continued)\\
(37) Honeycutt et al. (1995)\\
(38) Thorstensen \& Taylor (2000)\\
(39) Vanlandingham et al. (1996)\\
(40) Matheson et al. (1993)\\
(41) Popper (1940)\\
(42) Larsson-Leander (1954)\\
(43) Diaz et al. (1995)\\
(44) P\'equignot et al. (1993)\\ 
(45) Morisset \& P\'equignot (1995)\\
(46) Zwitter \& Munari (1996)\\
(47) Lynch et al. (1989) \\
(48) Weiler \& Bahng (1976)\\
(49) Payne-Gaposchkin \& Menzel (1938)\\
(50) Zwitter \& Munari (1995)\\
(51) Saizar et al. (1996)\\
(52) Scott et al. (1995)\\
(53) Kawabata et al. (2000)\\
(54) Ciatti \& Rosino (1974)\\
(55) Anupama et al. (1992)\\
(56) Szkody \& Howell (1992)\\
(57) Della Valle et al. (1997)\\
(58) Saizar et al. (1991)\\
(59) Rosino et al. (1992)\\
(60) Saizar et al. (1992)\\
(61) Scott et al. (1994)\\
(62) Gehrz et al. (1992)\\
(63) Hachisu \& Kato (2000a)\\
(64) Hachisu \& Kato (2000b)\\
(65) Tolbert et al. (1967)\\
(66) Bohigas et al. (1989)\\
(67) Gilmozzi et al. (1998)\\
(68) Gonz\'alez-Riestra (1992)\\
(69) Anupama \& Sethi (1994)\\
(70) Hachisu et al. (2000)\\
(71) Anupama \& Dewangan (2000)\\
(72) Sekiguchi et al. (1990a)\\
(73) Ciardullo et al. (1987)\\
(74) Ciardullo et al. (1990)\\
(75) Tomaney \& Shafter (1993)\\
(76) Shore et al. (1991)\\
(77) Sekiguchi et al. (1990b)\\

\clearpage

\begin{table}
\begin{center}
\footnotesize

Table 2. \OIII Flux and Luminosity
\vspace*{1mm}

\begin{tabular}{lrrrr}\hline\hline
Object & Age & Flux\phantom{ergm} & Ref. & Luminosity\\
       & (years) & (erg cm$^{-2}$ s$^{-1}$) & & (erg s$^{-1}$) \\ \hline
OS And~~~ &   4.6 & $3.4 \times 10^{-14}$ & (2) & $2.5 \times 10^{32}$ \\
DO Aql    &  74.9 & $9.3 \times 10^{-15}$ & (4) & $7.3 \times 10^{31}$ \\
V603 Aql  &  59.8$^{\rm a}$ & $2.1 \times 10^{-14}$ & (6) & $3.5 \times 10^{29}$ \\
          &  0.89 & $4.9 \times 10^{-7}$  & (7) & $8.2 \times 10^{36}$ \\ 
          &  0.46 & $4.7 \times 10^{-7}$  & (7) & $7.9 \times 10^{36}$ \\ 
          &  0.33 & $3.5 \times 10^{-7}$  & (7) & $5.9 \times 10^{36}$ \\
\hline
\end{tabular}
\end{center}
\end{table}

\clearpage

\begin{table}
\begin{center}
\footnotesize

Table 2. \OIII Flux and Luminosity (continued)
\vspace*{1mm}

\begin{tabular}{lrrrr}\hline\hline
Object & Age & Flux\phantom{ergm} & Ref. & Luminosity\\
       & (years) & (erg cm$^{-2}$ s$^{-1}$) & & (erg s$^{-1}$) \\ \hline
V888 Cen  &   3.1 & $2.1 \times 10^{-13}$ & (4) & $ 1.0\times 10^{34}$\\
          &   2.8 & $2.6 \times 10^{-12}$ & (24) & $ 1.3\times 10^{34}$\\
IV Cep    &  13.1 & absent                & (4) & \\
          &   8.3$^{\rm a}$ & $3.9 \times 10^{-16}$ & (6)  & $ 1.1\times 10^{31}$\\
          &  0.31 & $1.6 \times 10^{-11}$ & (25) & $4.4 \times 10^{35}$ \\
          &  0.21 & $1.6 \times 10^{-11}$ & (25) & $4.4 \times 10^{35}$ \\
\hline
\end{tabular}
\end{center}
\end{table}

\clearpage

\begin{table}
\begin{center}
\footnotesize

Table 2. \OIII Flux and Luminosity (continued)
\vspace*{1mm}

\begin{tabular}{lrrrr}\hline\hline
Object & Age & Flux\phantom{ergm} & Ref. & Luminosity\\
       & (years) & (erg cm$^{-2}$ s$^{-1}$) & & (erg s$^{-1}$) \\ \hline
DN Gem    &   2.9 & absent                & (26) \\
          &   1.9 & present               & (26) \\
DQ Her    &  60.4$^{\rm a}$ & $5.5 \times 10^{-15}$ & (6) & $2.2 \times 10^{29}$ \\
          &  49.8 & $1.2 \times 10^{-13}$ & (3) & $4.6 \times 10^{30}$ \\
          &  49.7 & $7.6 \times 10^{-14}$ & (4) & $4.0 \times 10^{30}$ \\
          &  43.4$^{\rm a}$ & $4.3 \times 10^{-15}$ & (6) & $1.7 \times 10^{29}$ \\
\hline
\end{tabular}
\end{center}
\end{table}

\clearpage

\begin{table}
\begin{center}
\footnotesize

Table 2. \OIII Flux and Luminosity (continued)
\vspace*{1mm}

\begin{tabular}{lrrrr}\hline\hline
Object & Age & Flux\phantom{ergm} & Ref. & Luminosity\\
       & (years) & (erg cm$^{-2}$ s$^{-1}$) & & (erg s$^{-1}$) \\ \hline
GQ Mus    &   2.3 & $1.6 \times 10^{-11}$ & (45) & $1.8 \times 10^{35}$ \\
          &   1.4 & $1.3 \times 10^{-10}$ & (45) & $1.5 \times 10^{36}$ \\
          &   1.2 & $1.5 \times 10^{-10}$ & (45) & $1.7 \times 10^{36}$ \\
V841 Oph  &  130.$^{\rm a}$ & $9.7 \times 10^{-15}$ &  (6) \\
V972 Oph  &  42.8 & absent                &  (4) \\
          &  37.6 & absent                & (46) \\
\hline
\end{tabular}
\end{center}
\end{table}

\clearpage

\begin{table}
\begin{center}
\footnotesize

Table 2. \OIII Flux and Luminosity (continued)
\vspace*{1mm}

\begin{tabular}{lrrrr}\hline\hline
Object & Age & Flux\phantom{ergm} & Ref. & Luminosity\\
       & (years) & (erg cm$^{-2}$ s$^{-1}$) & & (erg s$^{-1}$) \\ \hline
RR Pic    &  1.49 & $5.6 \times 10^{-9}$  & (49) & $2.7 \times 10^{35}$ \\
          &  1.38 & $4.2 \times 10^{-9}$  & (49) & $2.0 \times 10^{35}$ \\
          &  1.32 & $5.9 \times 10^{-9}$  & (49) & $2.9 \times 10^{35}$ \\
          &  1.25 & $4.3 \times 10^{-9}$  & (49) & $2.1 \times 10^{35}$ \\
          &  1.15 & $3.1 \times 10^{-9}$  & (49) & $1.5 \times 10^{35}$ \\
          &  0.99 & $4.9 \times 10^{-9}$  & (49) & $2.4 \times 10^{35}$ \\
\hline
\end{tabular}
\end{center}
\end{table}

\clearpage

\begin{table}
\begin{center}
\footnotesize

Table 2. \OIII Flux and Luminosity (continued)
\vspace*{1mm}

\begin{tabular}{lrrrr}\hline\hline
Object & Age & Flux\phantom{ergm} & Ref. & Luminosity\\
       & (years) & (erg cm$^{-2}$ s$^{-1}$) & & (erg s$^{-1}$) \\ \hline
V977 Sco  & 1.819 &  $5.0 \times 10^{-14}$ & (13) & $7.7 \times 10^{34}$ \\
          & 1.000 &  $1.4 \times 10^{-12}$ & (13) & $2.2 \times 10^{36}$ \\
          & 0.805 &  $7.8 \times 10^{-13}$ & (13) & $1.2 \times 10^{36}$ \\
          & 0.750 &  $7.4 \times 10^{-13}$ & (13) & $1.1 \times 10^{36}$ \\
          & 0.666 &  $9.0 \times 10^{-13}$ & (13) & $1.4 \times 10^{36}$ \\
          & 0.630 &  $8.0 \times 10^{-13}$ & (13) & $1.2 \times 10^{36}$ \\
\hline
\end{tabular}
\end{center}
\end{table}

\clearpage

\begin{table}
\begin{center}
\footnotesize

Table 2. \OIII Flux and Luminosity (continued)
\vspace*{1mm}

\begin{tabular}{lrrrr}\hline\hline
Object & Age & Flux\phantom{ergm} & Ref. & Luminosity\\
       & (years) & (erg cm$^{-2}$ s$^{-1}$) & & (erg s$^{-1}$) \\ \hline
LW Ser    &  22.4 & $2.3 \times 10^{-15}$  &  (4) \\   
XX Tau    &  56.9 & absent                 &  (4) \\
RW UMi    &  41.7 & absent                 &  (4) \\
          &  27.9 & absent                 &  (4) \\ 
LV Vul    &  23.3 & absent                 &  (2) \\
NQ Vul    &  21.6$^{\rm a}$ 
                  & $9.4 \times 10^{-16}$  &  (6) & $1.4 \times 10^{30}$ \\
\hline
\end{tabular}
\end{center}
\end{table}

\clearpage

\begin{table}
\begin{center}
\footnotesize

Table 2. \OIII Flux and Luminosity (continued)
\vspace*{1mm}

\begin{tabular}{lrrrr}\hline\hline
Object & Age & Flux\phantom{ergm} & Ref. & Luminosity\\
       & (years) & (erg cm$^{-2}$ s$^{-1}$) & & (erg s$^{-1}$) \\ \hline
V3890 Sgr & 0.093 & absent                 & (13)                        \\
          & 0.077 & absent                 & (13)                        \\
          & 0.049 & absent                 & (69)                        \\
          & 0.024 & absent                 & (13)                        \\
U Sco     & 0.032 & absent                 & (71)                        \\
V745 Sco  & 2.679 & absent                 & (13)                        \\
\hline
\end{tabular}

$^{\rm a}$spectrum with no emission present

\end{center}
\vspace*{10cm}
\end{table}

\clearpage

\begin{table}
\begin{center}
\footnotesize

Table 3. H$\alpha$ Flux and Luminosity
\vspace*{1mm}

\begin{tabular}{lrrrr}\hline\hline
Object & Age & Flux\phantom{ergm} & Ref. & Luminosity\\
       & (years) & (erg cm$^{-2}$ s$^{-1}$) & & (erg s$^{-1}$) \\ \hline
OS And~~~ &   4.6 & $8.4 \times 10^{-14}$ &  (2) & $4.7 \times 10^{32}$ \\
DO Aql    &  74.9 & $1.5 \times 10^{-14}$ &  (4) & $1.0 \times 10^{32}$ \\
V603 Aql  &  78.0 & $5.7 \times 10^{-13}$ &  (6)?? & $8.9 \times 10^{30}$ \\
          &  76.9 & $5.5 \times 10^{-13}$ &  (6)?? & $8.6 \times 10^{30}$ \\
          &  72.2 & $7.0 \times 10^{-13}$ &  (2) & $1.1 \times 10^{31}$ \\
V1370 Aql &   9.5 & $6.7 \times 10^{-15}$ &  (2) & $8.1 \times 10^{31}$ \\
\hline
\end{tabular}
\end{center}
\end{table}

\clearpage

\begin{table}
\begin{center}
\footnotesize

Table 3. H$\alpha$ Flux and Luminosity (continued)
\vspace*{1mm}

\begin{tabular}{lrrrr}\hline\hline
Object & Age & Flux\phantom{ergm} & Ref. & Luminosity\\
       & (years) & (erg cm$^{-2}$ s$^{-1}$) & & (erg s$^{-1}$) \\ \hline
V1500 Cyg &  0.05 & $1.4 \times 10^{-08}$ & (28) & $1.2 \times 10^{37}$ \\
          &  0.01 & $5.9 \times 10^{-08}$ & (28) & $5.2 \times 10^{37}$ \\
V1819 Cyg &  11.8 & $8.8 \times 10^{-15}$ &  (4) & $1.3 \times 10^{32}$ \\
          &  1.8  & $8.9 \times 10^{-13}$ & (29) & $1.3 \times 10^{34}$ \\
          &  0.9  & $8.2 \times 10^{-11}$ & (29) & $2.2 \times 10^{36}$ \\
          &  0.3  & $6.6 \times 10^{-11}$ & (29) & $9.8 \times 10^{35}$ \\
\hline
\end{tabular}
\end{center}
\end{table}

\clearpage

\begin{table}
\begin{center}
\footnotesize

Table 3. H$\alpha$ Flux and Luminosity (continued)
\vspace*{1mm}

\begin{tabular}{lrrrr}\hline\hline
Object & Age & Flux\phantom{ergm} & Ref. & Luminosity\\
       & (years) & (erg cm$^{-2}$ s$^{-1}$) & & (erg s$^{-1}$) \\ \hline
CP Lac    & 0.05  & $2.8 \times 10^{-8}$  & (41) & $1.0 \times 10^{37}$ \\
          & 0.04  & $3.3 \times 10^{-8}$  & (41) & $1.2 \times 10^{37}$ \\
DK Lac    &  41.4 & $7.6 \times 10^{-15}$ &  (2) & $3.9 \times 10^{31}$ \\
          &  0.24 &                       & (42) & $1.2 \times 10^{36}$ \\
          &  0.21 &                       & (42) & $3.7 \times 10^{35}$ \\
          &  0.16 &                       & (42) & $6.9 \times 10^{35}$ \\
\hline
\end{tabular}
\end{center}
\end{table}

\clearpage

\begin{table}
\begin{center}
\footnotesize

Table 3. H$\alpha$ Flux and Luminosity (continued)
\vspace*{1mm}

\begin{tabular}{lrrrr}\hline\hline
Object & Age & Flux & Ref. & Luminosity\\
       & (years) & (erg cm$^{-2}$ s$^{-1}$) & & (erg s$^{-1}$) \\ \hline
CP Pup    &  52.9 & $1.1 \times 10^{-13}$ &  (4) & $6.9 \times 10^{31}$ \\
V351 Pup  &   6.1 & $1.8 \times 10^{-14}$ &  (4) & $8.5 \times 10^{31}$ \\
          & 1.325 & $1.1 \times 10^{-11}$ & (13) & $5.2 \times 10^{34}$ \\
          &   1.3 & $1.1 \times 10^{-11}$ & (51) & $5.2 \times 10^{34}$ \\
          & 1.237 & $1.4 \times 10^{-11}$ & (13) & $6.6 \times 10^{34}$ \\
          &   1.2 & $1.5 \times 10^{-11}$ & (51) & $7.1 \times 10^{34}$ \\
\hline
\end{tabular}
\end{center}
\end{table}

\clearpage

\begin{table}
\begin{center}
\footnotesize

Table 3. H$\alpha$ Flux and Luminosity (continued)
\vspace*{1mm}

\begin{tabular}{lrrrr}\hline\hline
Object & Age & Flux\phantom{ergm} & Ref. & Luminosity\\
       & (years) & (erg cm$^{-2}$ s$^{-1}$) & & (erg s$^{-1}$) \\ \hline
V992 Sco  & 0.186 & $2.6 \times 10^{-10}$ & (13) & $4.2 \times 10^{35}$ \\
          & 0.181 & $3.2 \times 10^{-10}$ & (13) & $5.2 \times 10^{35}$ \\
          & 0.175 & $4.2 \times 10^{-10}$ & (13) & $6.8 \times 10^{35}$ \\
          & 0.164 & $5.3 \times 10^{-10}$ & (13) & $8.6 \times 10^{35}$ \\
          & 0.159 & $6.7 \times 10^{-10}$ & (13) & $1.1 \times 10^{36}$ \\
          & 0.153 & $6.2 \times 10^{-10}$ & (13) & $1.0 \times 10^{36}$ \\
\hline
\end{tabular}
\end{center}
\end{table}

\clearpage

\begin{table}
\begin{center}
\footnotesize

Table 3. H$\alpha$ Flux and Luminosity (continued)
\vspace*{1mm}

\begin{tabular}{lrrrr}\hline\hline
Object & Age & Flux\phantom{ergm} & Ref. & Luminosity\\
       & (years) & (erg cm$^{-2}$ s$^{-1}$) & & (erg s$^{-1}$) \\ \hline
QU Vul    &   5.7 & $1.3 \times 10^{-12}$ & (59) & $1.9 \times 10^{33}$ \\
          &   3.7 & $3.4 \times 10^{-12}$ & (59) & $5.1 \times 10^{33}$ \\
          &   2.6 & $8.0 \times 10^{-12}$ & (59) & $1.2 \times 10^{34}$ \\
QV Vul    &  10.5 & $5.4 \times 10^{-14}$ &  (4) & $1.2 \times 10^{32}$ \\
          & 5.71  & $4.1 \times 10^{-13}$ & (61) & $9.1 \times 10^{32}$ \\
          & 2.79  & $9.1 \times 10^{-12}$ & (62) & $2.0 \times 10^{34}$ \\
\hline
\end{tabular}
\end{center}
\end{table}

\clearpage

\begin{table}
\begin{center}
\footnotesize

Table 3. H$\alpha$ Flux and Luminosity (continued)
\vspace*{1mm}

\begin{tabular}{lrrrr}\hline\hline
Object & Age & Flux\phantom{1mm} & Ref. & Luminosity\\
       & (years) & (erg cm$^{-2}$ s$^{-1}$) & & (erg s$^{-1}$) \\ \hline
1991      & 0.003 & $3.1 \times 10^{-11}$ & (13) & $2.1 \times 10^{37}$ \\
1992      & 0.356 & $3.2 \times 10^{-13}$ & (13) & $2.1 \times 10^{35}$ \\
          & 0.214 & $2.6 \times 10^{-12}$ & (13) & $1.7 \times 10^{36}$ \\
          & 0.131 & $5.3 \times 10^{-12}$ & (13) & $3.5 \times 10^{36}$ \\
          & 0.096 & $1.4 \times 10^{-11}$ & (13) & $9.2 \times 10^{36}$ \\
          & 0.052 & $2.6 \times 10^{-11}$ & (13) & $1.7 \times 10^{37}$ \\
\hline
\end{tabular}

$^{\rm a}$flux includes [N II] lines

\end{center}
\vspace*{10cm}
\end{table}

\clearpage

\begin{table}
\begin{center}
\footnotesize

Table 4. H$\beta$ Flux and Luminosity
\vspace*{1mm}

\begin{tabular}{lrrrr}\hline\hline
Object & Age & Flux\phantom{ergm} & Ref. & Luminosity\\
       & (years) & (erg cm$^{-2}$ s$^{-1}$) & & (erg s$^{-1}$) \\ \hline
OS And~~~ &   4.6 & $6.0 \times 10^{-15}$ &  (2) & $4.3 \times 10^{31}$ \\
V603 Aql  &  78.0 & $5.6 \times 10^{-13}$ &  (6) & $9.3 \times 10^{30}$ \\
          &  72.2 & $6.0 \times 10^{-13}$ &  (2) & $1.0 \times 10^{31}$ \\
          &  59.8 & $1.4 \times 10^{-13}$ &  (6) & $2.3 \times 10^{30}$ \\
          &  0.89 & $7.8 \times 10^{-9}$  &  (7) & $1.3 \times 10^{35}$ \\
          &  0.46 & $2.8 \times 10^{-8}$  &  (7) & $4.7 \times 10^{35}$ \\
\hline
\end{tabular}
\end{center}
\end{table}

\clearpage

\begin{table}
\begin{center}
\footnotesize

Table 4. H$\beta$ Flux and Luminosity (continued)
\vspace*{1mm}

\begin{tabular}{lrrrr}\hline\hline
Object & Age & Flux\phantom{ergm} & Ref. & Luminosity\\
       & (years) & (erg cm$^{-2}$ s$^{-1}$) & & (erg s$^{-1}$) \\ \hline
V476 Cyg  &  0.15 & $2.3 \times 10^{-9}$  &  (7) & $1.7 \times 10^{36}$ \\
          &  0.13 & $4.4 \times 10^{-9}$  &  (7) & $3.3 \times 10^{36}$ \\
          &  0.12 & $4.7 \times 10^{-9}$  &  (7) & $3.6 \times 10^{36}$ \\
          &  0.11 & $8.1 \times 10^{-9}$  &  (7) & $6.1 \times 10^{36}$ \\
          &  0.09 & $1.0 \times 10^{-8}$  &  (7) & $7.6 \times 10^{36}$ \\
          &  0.08 & $1.6 \times 10^{-8}$  &  (7) & $1.2 \times 10^{37}$ \\
\hline
\end{tabular}
\end{center}
\end{table}

\clearpage

\begin{table}
\begin{center}
\footnotesize

Table 4. H$\beta$ Flux and Luminosity (continued)
\vspace*{1mm}

\begin{tabular}{lrrrr}\hline\hline
Object & Age & Flux & Ref. & Luminosity\\
       & (years) & (erg cm$^{-2}$ s$^{-1}$) & & (erg s$^{-1}$) \\ \hline
HR Del    &  2.0  & $6.5 \times 10^{-10}$ & (31) & $7.5 \times 10^{34}$ \\
          & 1.741 & $7.3 \times 10^{-11}$ & (3)  & $8.4 \times 10^{33}$ \\
          & 1.667 & $1.1 \times 10^{-10}$ & (3)  & $1.3 \times 10^{34}$ \\
          & 0.964 & $5.2 \times 10^{-10}$ & (3)  & $5.9 \times 10^{34}$ \\
          & 0.550 & $1.0 \times 10^{-9} $ & (3)  & $1.2 \times 10^{35}$ \\
          & 0.449 & $1.0 \times 10^{-9} $ & (3)  & $1.2 \times 10^{35}$ \\
\hline
\end{tabular}
\end{center}
\end{table}

\clearpage

\begin{table}
\begin{center}
\footnotesize

Table 4. H$\beta$ Flux and Luminosity (continued)

\begin{tabular}{lrrrr}\hline\hline
Object & Age & Flux\phantom{ergm} & Ref. & Luminosity\\
       & (years) & (erg cm$^{-2}$ s$^{-1}$) & & (erg s$^{-1}$) \\ \hline
GQ Mus    &   3.1 & $1.1 \times 10^{-12}$ & (45) & $1.2 \times 10^{34}$ \\
          &   3.0 & $1.4 \times 10^{-12}$ & (45) & $1.6 \times 10^{34}$ \\
          &   2.3 & $6.8 \times 10^{-12}$ & (45) & $7.6 \times 10^{34}$ \\
          &   2.3 & $7.6 \times 10^{-12}$ & (45) & $8.5 \times 10^{34}$ \\
          &   1.5 & $5.1 \times 10^{-11}$ & (45) & $5.7 \times 10^{35}$ \\
          &   1.4 & $6.3 \times 10^{-11}$ & (45) & $7.1 \times 10^{35}$ \\
\hline
\end{tabular}
\end{center}
\end{table}

\clearpage

\begin{table}
\begin{center}
\footnotesize

Table 4. H$\beta$ Flux and Luminosity (continued)
\vspace*{1mm}

\begin{tabular}{lrrrr}\hline\hline
Object & Age & Flux\phantom{ergm} & Ref. & Luminosity\\
       & (years) & (erg cm$^{-2}$ s$^{-1}$) & & (erg s$^{-1}$) \\ \hline
RR Pic    &  3.66 & $7.7 \times 10^{-10}$ & (49) & $3.7 \times 10^{34}$ \\
          &  2.86 & $2.5 \times 10^{-9}$  & (49) & $1.2 \times 10^{35}$ \\
          &  2.56 & $2.8 \times 10^{-9}$  & (49) & $1.4 \times 10^{35}$ \\
          &  1.87 & $3.6 \times 10^{-9}$  & (49) & $1.8 \times 10^{35}$ \\
          &  1.79 & $4.0 \times 10^{-9}$  & (49) & $1.9 \times 10^{35}$ \\
          &  1.75 & $2.2 \times 10^{-9}$  & (49) & $1.1 \times 10^{35}$ \\
\hline
\end{tabular}
\end{center}
\end{table}

\clearpage

\begin{table}
\begin{center}
\footnotesize

Table 4. H$\beta$ Flux and Luminosity (continued)

\begin{tabular}{lrrrr}\hline\hline
Object & Age & Flux\phantom{ergm} & Ref. & Luminosity\\
       & (years) & (erg cm$^{-2}$ s$^{-1}$) & & (erg s$^{-1}$) \\ \hline
V4169 Sgr & 0.700 & $4.0 \times 10^{-13}$ & (13) & $6.7 \times 10^{33}$ \\
          & 0.295 & $7.0 \times 10^{-12}$ & (13) & $1.2 \times 10^{35}$ \\
          & 0.053 & $5.1 \times 10^{-11}$ & (13) & $8.6 \times 10^{35}$ \\
V4171 Sgr & 0.701 & $8.4 \times 10^{-14}$ & (13) & $1.4 \times 10^{33}$ \\
          & 0.529 & $2.7 \times 10^{-13}$ & (13) & $4.3 \times 10^{33}$ \\
          & 0.441 & $4.4 \times 10^{-13}$ & (13) & $7.1 \times 10^{33}$ \\
\hline
\end{tabular}
\end{center}
\end{table}

\clearpage

\begin{table}
\begin{center}
\footnotesize

Table 4. H$\beta$ Flux and Luminosity (continued)
\vspace*{1mm}

\begin{tabular}{lrrrr}\hline\hline
Object & Age & Flux\phantom{ergm} & Ref. & Luminosity\\
       & (years) & (erg cm$^{-2}$ s$^{-1}$) & & (erg s$^{-1}$) \\ \hline
V444 Sct  & 0.186 & $5.7 \times 10^{-14}$ & (13) & $2.0 \times 10^{34}$ \\
          & 0.110 & $2.8 \times 10^{-13}$ & (13) & $9.7 \times 10^{34}$ \\
          & 0.006 & $1.2 \times 10^{-11}$ & (13) & $4.1 \times 10^{35}$ \\
X Ser     &  96.0 & $5.4 \times 10^{-15}$ & (38) & $1.4 \times 10^{31}$ \\
          &  94.0 & $4.5 \times 10^{-15}$ & (38) & $1.1 \times 10^{31}$ \\
          &  86.9 & $8.0 \times 10^{-15}$ & (55) & $2.0 \times 10^{31}$ \\
\hline
\end{tabular}
\end{center}
\end{table}

\clearpage

\begin{table}
\begin{center}
\footnotesize

Table 4. H$\beta$ Flux and Luminosity (continued)
\vspace*{1mm}

\begin{tabular}{lrrrr}\hline\hline
Object & Age & Flux & Ref. & Luminosity\\
       & (years) & (erg cm$^{-2}$ s$^{-1}$) & & (erg s$^{-1}$) \\ \hline
RS Oph    & 0.020 & $6.4 \times 10^{-10}$ & (67) & $2.8 \times 10^{35}$ \\
          & 0.015 & $1.0 \times 10^{-9}$  & (67) & $4.4 \times 10^{35}$ \\
          & 0.007 & $2.6 \times 10^{-9}$  & (67) & $1.1 \times 10^{36}$ \\
          & 0.004 & $3.3 \times 10^{-9}$  & (67) & $1.4 \times 10^{36}$ \\
          & 0.001 & $2.6 \times 10^{-9}$  & (67) & $1.1 \times 10^{36}$ \\
V3890 Sgr & 1.937 & $1.8 \times 10^{-14}$ & (13) & $1.1 \times 10^{33}$ \\
\hline
\end{tabular}
\end{center}
\vspace*{10cm}
\end{table}

\end{document}